\documentclass[aps,prb,groupaddress,reprint,twocolomn]{revtex4-1}
\usepackage{amssymb}
\usepackage[pdftex]{graphicx}
\usepackage{amsmath}
\usepackage{amsfonts}
\usepackage{color}
\usepackage{here}
\usepackage{bm}%
\usepackage{ulem}

\newcommand{\Uline}[1]{\textcolor{black}{#1}}

\begin{document}
\title{Statistical and Analytical Approaches\\ to
  Finite Temperature Magnetic Properties of SmFe$_{12}$ compound}
\date{}
\author{Takuya Yoshioka}
\affiliation{Department of Applied Physics, Tohoku University, Sendai 980-8579, Japan}
\affiliation{ESICMM, National Institute for Materials Science, Tsukuba, Ibaraki 305-0047, Japan}
\affiliation{Center for Spintronics Research Network, Tohoku University, Sendai 980-8577, Japan}
\author{Hiroki Tsuchiura}
\affiliation{Department of Applied Physics, Tohoku University, Sendai 980-8579, Japan}
\affiliation{ESICMM, National Institute for Materials Science, Tsukuba, Ibaraki 305-0047, Japan}
\affiliation{Center for Spintronics Research Network, Tohoku University, Sendai 980-8577, Japan}
\author{Pavel Nov\'ak}
\affiliation{Institute of Physics of ASCR, Cukrovarnick\'a, Prague 6 162 00, Czech Republic}

\begin{abstract}
To investigate the magnetic properties of SmFe$_{12}$, we construct an
effective spin model, where magnetic moments, crystal field (CF) parameters, and
exchange fields at 0 K are determined by first principles. Finite temperature
magnetic properties are investigated by using this model.
We further develop an analytical method with strong
mixing of states with different quantum number of angular momentum $J$ ($J$-mixing),
which is caused by strong exchange field acting on spin component of $4f$ electrons.
Comparing our
analytical results with those calculated by Boltzmann statistics,
we clarify that the previous analytical studies for Sm transition metal compounds
over-estimate the $J$-mixing effects.
The present method
enables us to make quantitative analysis of temperature dependence of magnetic
anisotropy (MA) with high-reliability.
The analytical method with model approximations reveals that
  the $J$-mixing caused by exchange field increases spin angular momentum,
  which enhances the absolute value of orbital angular momentum and MA constants
via spin-orbit interaction.
It is also clarified that
these $J$-mixing effects remain even above room temperature.
Magnetization of SmFe$_{12}$ shows
peculiar field dependence known as first-order magnetization process
(FOMP), where the magnetization shows an abrupt change at certain magnetic field.
The result of the analysis shows that the origin of FOMP is attributed
to competitive MA constants between positive $K_{1}$ and negative $K_2$.
The sign of $K_{1(2)}$ appears due to an increase in
CF potential denoted by the parameter $A_{2}^{0}\langle r^{2}\rangle$ ($A_4^0\langle r^4\rangle$)
  caused by hybridization between $3d$-electrons of Fe on $8i$ ($8j$) site and
  $5d$ and $6p$ valence electrons on Sm site.
It is verified that the requirement for the appearance of FOMP is given as
$-K_2<K_{1}<-6K_{2}$.

\end{abstract}
\pacs{Valid PACS appear here}% PACS, the Physics and Astronomy
                             % Classification Scheme.
%¥keywords{Suggested keywords}%Use showkeys class option if keyword
                              %display desired
\maketitle

\section{Introduction}

\begin{figure}[htbp]
\begin{center}
\includegraphics[width=6.5cm]{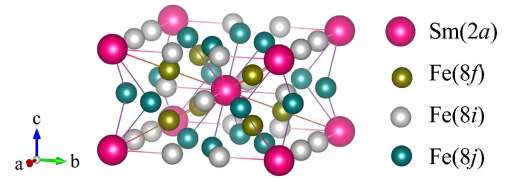}
\end{center}
\caption{(color online) Crystal structure of SmFe$_{12}$ compound
    in ThMn$_{12}$ structure.
Inequivalent sites: Sm(2$a$), Fe(8$f$), Fe(8$i$), and Fe(8$j$) are
shown by different-colored balls
and solid lines show interatomic short contacts less than 3.2 \AA.}%
\label{fig:struct}%
\end{figure}

There have been intensive studies on developing new rare-earth ($R$) lean
permanent magnetic materials which have strong magnetic properties comparable
to those of Nd-Fe-B. Nitrogenated compounds as NdFe$_{12}$N or NdFe$_{11}$TiN
have been considered to be candidates of such materials,
and thus series of experimental and
theoretical efforts have been made to figure out the magnetic properties of
these materials \cite{Miyake,Hirayama1}. SmFe$_{12}$ with the ThMn$_{12}$
structure (Fig. \ref{fig:struct}) is also a possible candidate and has attracted renewed interest
because it exhibits excellent intrinsic magnetic properties such as uniaxial
magnetocrystalline anisotropy \cite{Hadjipanayis}. Although SmFe$_{12}$ itself is
thermodynamically unstable, it has been known that the substitution of Fe with
a stabilizing element, such as Ti or V, can remove this difficulty
\cite{Ohashi1,Ohashi2,Hu,Kuno,Schoenhoebel}. In these systems, however, the saturated
magnetization is reduced due to anti-parallel alignment of magnetic moments of
Ti and V relative to
those of Fe. Recent development of the synthesis technology made
it possible to fabricate highly textured single phase samples of SmFe$_{12}$
thin film \cite{Kato,Hirayama2,Ogawa1,Ogawa2,Sepehri-Amin3}, and it has been shown experimentally
that Co substitution for Fe enhances their magnetic properties, such as Curie
temperature and magnetic anisotropy (MA) \cite{Hirayama2}. Thus,
SmFe$_{12}$-based systems belong to
one of the most promising hard magnetic materials, and
therefore to clarify the basic magnetic properties of SmFe$_{12}$ is crucially
important.
%%%%%%%%%%%%%%%%%%%%%%%%%%%%%%%%%%%%%%%%%%%%%%%%%%%%%%
%%%%%%%%%%%%%%%%%%%%%%%%%%%%%%%%%%%%%%%%%%%%%%%%%%%%%%%

So far many attempts have been performed for microscopic
understanding of the magnetic properties of $R$ based permanent magnets
\cite{Evans, Toga, Nishino, MiyakeAkai, Yoshioka, Tsuchiura2}. Among them a
powerful method is to combine the first-principle calculations for electronic
states at the ground state with a suitable model for finite temperature
properties
\cite{Yamada,Sasaki,Miura-direct,Miura-pl,Kuzminlinear,Kuzmin2nd,Kuzminlimit,Millev,Kuzminmix,Magnani}%
. As for SmFe$_{12}$, Harashima \textit{et al}. (2015) \cite{Harashima},
K\"{o}rner \textit{et al} (2016) \cite{Koerner} and Delonge \textit{et al}
(2017) \cite{Delange} performed the first-principle calculations and model
analysis of
magnetic properties. In the theoretical study of Sm-based
intermetallic compounds, however, there remains a basic issue how to deal with
the formidably strong $J$-mixing effects in Sm. This is the problem studied
for a long period on the Sm-based magnets \cite{VanVleck,Sankar,Wijin}. There
are some attempts to include the $J$-mixing in the analytical form by the
first-order perturbation for the crystal fields (CFs)
\cite{Kuzminmix,Magnani}. However, Kuz'min pointed out that the Sm-based
magnetic materials are exceptional for application of the method
\cite{Kuzminmix}.

We have recently developed a similar method \cite{Yoshioka,Tsuchiura2}, in
which the model parameters are calculated by the first-principles and the
finite temperature magnetic properties are calculated in a statistical way,
and applied it to $R_{2}$Fe$_{14}$B systems. By taking into account
CF parameters up to 6-th order,
the model satisfactorily explained the experimental
results for magnetization curves and the temperature dependence of MA
constants \cite{Yoshioka,Tsuchiura2}. Using the method we recently calculated
the temperature dependence of the MA constants of SmFe$_{12}$ and showed that
$K_{1}>0$ and $K_{2}<0$ in consistence
with experimental results \cite{Ogawa2}. The report of the work,
however, contains only the final
results and no details of computational procedure have been presented. As a
results no explanations on the mechanism for the results that $K_{1}>0$ and
$K_{2}<0$ have been given.

The purpose of the present study is thus to clarify the origin of
the finite temperature magnetic properties
of SmFe$_{12}$ compound by statistical and analytical ways.
To this end, we describe the
details of the statistical method and develop a novel analytical method.
%The purpose of the present study is thus twofold; one is to describe the
%details of the method and results for the numerical method, and the other is
%to develop a novel analytical method.
%The origin of magnetic properties at
%finite temperature is clarified complementary by these methods.
The analytical procedure is able to derive simple relations
between the temperature dependence of
magnetic properties and parameters determined
by first-principles electronic structure calculations.
The treatment of the $J$-mixing effects adopted previously by the
other groups \cite{Kuzminmix,Magnani} will be modified,
and the results will be
compared with the statistical results of the temperature dependence of magnetic properties of SmFe$_{12}$.
Good agreement between the analytical and statistical results guarantees
the applicability of the modified analytical formula to Sm compounds.

In the following, we give the model Hamiltonian, the parameters
of which are determined by the
first-principles, and present the calculation procedure for
finite temperatures, especially the statistical method
to obtain the MA constants and magnetization curves,
and explain the modified analytical method.
The latter method may clarify the relations among the free energy of the system, the
CF, and the exchange field. Using the analytical method, we will
show that the mechanism of $K_{1}>0$ and $K_{2}<0$ in SmFe$_{12}$ is
attributed to the characteristic lattice structure around Sm ions, that is,
crystallographic 2$b$-sites on $c$-axis adjacent to Sm are vacant. We also present
results on the magnetization process and nucleation fields by calculating
Gibbs free energy. As pointed out in
Ref. \cite{Tsuchiura2}, this analytical
spin model can be easily extended to Sm ions around the intergranular phases,
which is crucially important in the coercivity mechanism
\cite{Sepehri-Amin1,Sepehri-Amin2,Sepehri-Amin3}.

This paper is organized as follows.
The model Hamiltonian is explained in Section II,
and the procedure of the statistical and analytical method are explained in Section III.
Section IV shows the results of temperature dependence of magnetic properties calculated
in the statistical and analytical methods.
A summary of our work is given in Section V.

\section{Model Hamiltonian}

We adopt a following Hamiltonian to investigate the magnetic properties of
$R$ transition metal ($TM$) compounds:
\begin{align}
  \hat{\cal H}=&\frac{1}{V_0}\sum_{j=1}^{n_R}\hat{\cal H}_{R,j}\nonumber\\
  &+K_{1}^{TM}(T)\sin^{2}\theta^{TM}-{\bm M}^{TM}(T)\cdot{\bm B},\label{eq:Htot}
\end{align}  
where $\hat{\cal H}_{R,j}$ is a Hamiltonian for $R$ ion on $j$-th site
and \Uline{$n_R$ is the number of $R$ ion} in the unit cell volume $V_0$.
Second and third term represent the phenomenological treatment of MA energy and
Zeeman term on $TM$ sublattice,
where $K_{1}^{TM}(T)$ and ${\bm M}^{TM}(T)$
are the temperature dependent anisotropy constant and magnetization
vector of $TM$ sublattice, respectively, and $\theta
^{TM}$ is the polar angle of ${\bm M}^{TM}(T)$ against the $c$-axis.
${\bm M}^{TM}(T)$ is given as $M^{TM}(T){\bm e}^{TM}$ by using the
absolute value of the sublattice magnetization $M^{TM}(T)$ and a
directional vector ${\bm e}^{TM}$ of ${\bm M}^{TM}(T)$. $M^{TM}(T)$
is defined by a part of magnetization subtracting the $4f$ electron
contribution from the total magnetization. ${\bm B}$ is an applied field.

\subsection{Hamiltonian of Single $R$ Ion}

The Hamiltonian for $4f$ shell in $j$-th $R$ ion in Eq. (\ref{eq:Htot}) is 
\begin{align}
  \hat{\cal H}_{R,j}&=\sum_{i=1}^{n_{4f}}\hat{h}_j(i)+\frac{1}{8\pi\varepsilon_0}
  \sum_{i\ne i'=1}^{n_{4f}}\frac{e^2}{|\hat{\bm r}_i-\hat{\bm r}_{i'}|},
  \label{eq:HRC}
\end{align}
with 
\begin{align}
\hat{h}_j(i)
=&  \xi \hat{\bm l}_i\cdot\hat{\bm s}_i+2\mu_B
   \hat{\bm s}_i\cdot{\bm B}_{{\rm ex},j}(T)\nonumber\\
&+\int r_i^2|R_{4f}(r_i)|^2V_j({\bm r}_i) dr_i\nonumber\\
&+\mu_B(\hat{\bm l}_i+2\hat{\bm s}_i)\cdot{\bm B}.
  \label{eq:Hsingle}
\end{align}
The first and second terms in Eq. (\ref{eq:HRC})
represent the single electron contribution
and the electron-electron repulsion in $4f$ shell, respectively,
where $n_{4f}$ is the number of $4f$ electrons,
$\varepsilon_0$ and $e$ are the vacuum permittivity and the elementary charge, respectively.
$\hat{h}_{j}(i)$ in Eq. (\ref{eq:Hsingle}) is
the Hamiltonian for $i$-th $4f$ electron on $j$-th $R$ site,
where the first term in Eq. (\ref{eq:Hsingle}) is the
spin-orbit interaction (SOI) between spin ($\hat{\bm s}_i$) and orbital ($\hat
{{\bm l}_i}$) angular momenta, with a coupling constant $\xi$.
The second term represents the exchange
interaction between spin moment and temperature dependent exchange field
${\bm B}_{{\rm ex},j}(T)=-{\bm e}^{TM}B_{{\rm ex},j}(T)$ on $j$-th $R$ site,
where $\mu_B$ is the Bohr magneton.
The third and fourth terms are the CF and Zeeman terms, respectively.
In the expression of CF,
$V_j({\bm r}_i)$ and $R_{4f}(r_i)$ are Coulomb potential and
radial parts of the $4f$ wave function on $j$-th $R$ site,
respectively.
Note that the kinetic energy and screened central potential terms are
effectively taken into account in the formation of $4f$ orbital.

To obtain the electronic properties at $T=0$,
we apply the first-principles and determine the parameters in the Hamiltonians
in Eq. (\ref{eq:Hsingle}). We use the full-potential linearized augmented plane
wave plus local orbitals (APW+lo) method implemented in the WIEN2k code
\cite{wien2k}. The Kohn-Sham equations are solved within the
generalized-gradient approximation (GGA). To simulate localized $4f$
states, we treat $4f$ states as atomic-like core states, which is so
called opencore method \cite{Richter1,Novak,Hummler0,Richter2,Divis1,Divis2}.

We calculate the ground state properties of SmFe$_{12}$ such as
Coulomb potential, charge distribution, and sublattice magnetizations.
In accord with the previous theoretical studies for SmFe$_{12}$
\cite{Harashima,Koerner,Delange}, we assume that Sm ion has
trivalent-like electronic structure.
The exchange fields $B_{{\rm ex},j}(0)$ at $T=0$ are determined from an
energy increase caused by spin flip of $4f$ electrons \cite{Brooks,Yoshioka},
and CFs acting on $i$-th $4f$
electron are directly estimated from Coulomb potential $V_j({\bm r}_i)$ acting on
$j$-th $R$ site. It is noted that 
the single ion Hamiltonian $\hat{\cal H}_{R,j}$
thus determined for $j$-th $R$ ions includes effects of $TM$
atoms surrounding the $R$ ions as a mean field. 

Practically, the CF term is rewritten as the following formula
\cite{Novak,Richter2}:
\begin{align}
  \int_0^{r_c} r_i^2|R_{4f}(r_i)|^2&V_j({\bm r}_i) dr_i=
  \sum_{l,m}\frac{A_{l,j}^{m}\langle r^l\rangle}{a_{l,m}}
  t_l^m(\hat{\theta}_i,\hat{\varphi}_i),\\
  A_{l,j}^{m}\langle r^{l}\rangle  =&a_{l,m}\int_{0}^{r_{\rm c}}dr_i%
  r_i^{2}|R_{4f}(r_i)|^{2}\nonumber\\
  &\times\int d\Omega_i
  V_j({\bm r}_i)t_l^m(\theta_i,\varphi_i),\label{eq:Alm}
\end{align}
where $A_{l,j}^{m}\langle r^{l}\rangle$ is
CF parameter on $j$-th $R$ site, $a_{l,m}$ is a numerical factor
\cite{Hutchings}, $t_l^m(\hat{\theta}_i,\hat{\varphi}_i)$ is
tesseral harmonic function of a solid angle $\Omega=(\hat{\theta}_i,\hat{\varphi}_i)$,
and $r_{\rm c}$ is a cut-off radius.

Values of CF parameters
$A_{l,j}^{m}\langle r^{l}\rangle$ in Eq. (\ref{eq:Alm}), exchange field
$B_{{\rm ex},j}(0)$ in Eq. (\ref{eq:Hsingle}), $TM$-sublattice
magnetization $M^{TM}(0)$ in Eq. (\ref{eq:Htot}) in SmFe$_{12}$
are shown in TABLE \ref{table:CFP}.
The lattice constants used in this calculations are the experimental values
$a=b=8.35$ \AA\ and $c=4.8$ \AA\ \cite{Hirayama2}.
For Wycoff positions, we apply the theoretically optimized ones
given in Ref. \cite{Harashima}.
The crystal structure of SmFe$_{12}$ is shown in Fig. \ref{fig:struct}.
%%%%%%%%%%%%%%%%%%%%%%%%%%%%%%%%%%%%%%%%%%%%%%%%%%%%%%
\begin{table*}[ptb]
\caption{Values of CF potentials $A_{l,j}^{m}\langle r^{l}\rangle$ [K], exchange
  field $\mu_{\rm B}B_{{\rm ex},j}(0)/k_{\rm B}$ [K], and
  $TM$-sublattice magnetization $V_0M^{TM}(0)$
  [$\mu_{\rm B}$] in SmFe$_{12}$ calculated by first-principles,
  where $\mu_{\rm B}$ and $k_{\rm B}$ are Bohr magneton and Boltzmann constant, respectively,
  and $V_0=a\times b\times c$.
We note that $A_{l,j}^{m}\langle r^{l}\rangle$ and
    $\mu_{\rm B}B_{{\rm ex},j}(0)/k_{\rm B}$
  are independent of site index $j$.}%
\begin{ruledtabular}
\begin{tabular}{ccccccc}
$A_{2,j}^0\langle r^2\rangle$&
$A_{4,j}^0\langle r^4\rangle$&
$A_{4,j}^4\langle r^4\rangle$&
$A_{6,j}^0\langle r^6\rangle$&
$A_{6,j}^4\langle r^6\rangle$&
$\mu_{\rm B}B_{{\rm ex},j}(0)/k_{\rm B}$&
  $V_0M^{TM}$(0)
  \\ \hline
-71.4  & -21.3 & -49.3 & 5.9 &3.0 & 296.1 & 51.6 
\end{tabular}
\end{ruledtabular}
\label{table:CFP}%
\end{table*}
%%%%%%%%%%%%%%%%%%%%%%%%%%%%%%%%%%%%%%%%%%%%%%%%%%%%%%%%%%%%

\subsection{Single $R$ Ion Hamiltonian \\in $LS$ Coupling Regime}

We here apply the concept of $LS$ coupling to
the single electron Hamiltonian of Eq. (\ref{eq:Hsingle})
with Russell Saunders states $|L,S;J,M\rangle$,
due to the strong Coulomb interaction between $4f$ electrons.
According to the Hund's rule,
we specify the quantum number of total orbital (spin) moment $L(S)$.
Total angular momentum $J$ is varied from $|L-S|$ to $L+S$,
and $M$ is the magnetic quantum number.
Thus the single ion Hamiltonian in Eq. (\ref{eq:HRC})
can be reduced to:
\begin{align}
\hat{\cal H}_R &  =\hat{\cal H}_{{\rm so}}+\hat
{\cal H}_{\rm ex}+\hat{\cal H}_{\rm CF}+\hat{\cal H}_{\rm Z},\label{eq:Heff}\\
\hat{\cal H}_{\rm so} &  =\lambda\hat{\bm L}\cdot
\hat{\bm S},\\
\hat{\cal H}_{\rm ex} &  =2\mu_{\rm B}\hat{\bm S}\cdot{\bm B}%
_{\rm ex}(T),\label{eq:Hex}\\
\hat{\cal H}_{\rm CF} & =   \sum_{l,m,m'}B_{l}^{m}\Theta
_{l}^{L}C_{m%
}^{(l)}(\hat{\bm L}),\label{HCF1}\\
\hat{\cal H}_{\rm Z} &  =\mu_{\rm B}(\hat{\bm L}+2\hat{\bm S})\cdot{\bm B}.%
\end{align}
Hereafter, the site index $j$ is omitted for single-ion quantity.
$\hat{\bm L}$ and $\hat{\bm S}$ are total orbital and spin momenta of $4f$ electrons, respectively,
 $B_{l}^{0}=\sqrt{(2l+1)/4\pi}A_{l}%
^{0}\langle r^{l}\rangle/a_{l,0}$ and $B_{l}^{\pm|m|}=(\mp1)^{m}%
\sqrt{(2l+1)/8\pi}\left[  A_{l}^{|m|}\langle r^{l}\rangle\mp iA_{l}%
  ^{-|m|}\langle r^{l}\rangle\right]  /a_{l,m}$ for $m\neq0$, and
$\Theta_l^{L}=\langle L\parallel \sum_i C^{(l)}_m(\hat{\theta}_i,
  \hat{\varphi}_i)\parallel L\rangle/\langle L\parallel
  \sum_i C^{(l)}_m(\hat{\bm L})\parallel L\rangle$.
In the treatment of SOI, we should note that the
eigenstates of $LS$ coupling are specified by the quantum number of $J$.
In general, the term $\hat{\cal H}_{\rm so}$ is dominating in Eq. (\ref{eq:Heff}).
  Thus $J$ is a good quantum number in most of the $R$-$4f$ systems.
 Because the $LS$ coupling in Sm compounds is weak compared with other $R$ ones,
 it is necessary to include excited $J$-multiplets.
Hereafter, we abbreviate the states $|L,S;J,M\rangle$ as $|J,M\rangle$.

%%%%%%%%%%%%%%%%%%%%%%%%%%%%%%%%%%%%%%%%%%%%%%%%%%%%%%%%%%%%%%%%%%%%%%%%%%%%%%
\begin{figure}[htbp]
\begin{center}
\includegraphics[width=8.0cm]{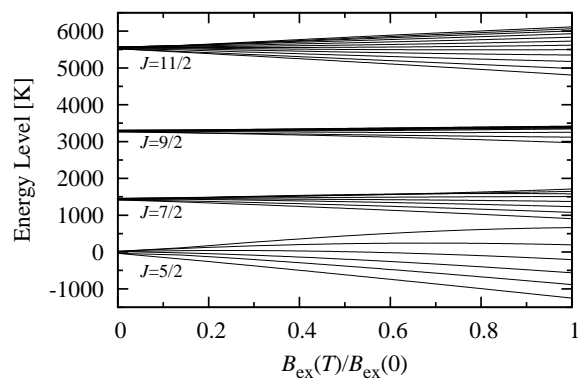}
\end{center}
\caption{Energy levels as a function of
  $B_{\rm ex}(T)/B_{\rm ex}(0)$ of the Sm-$4f$
  states in SmFe$_{12}$ at ${\bm e}^{TM} = {\bm n}_c$ and for ${\bm B}={\bm 0}$.
High energy levels originated from $J=13/2$ and $J=15/2$ multiplets are above 6000 K, which are not shown.}
\label{fig:engalpha}%
\end{figure}
%%%%%%%%%%%%%%%%%%%%%%%%%%%%%%%%%%%%%%%%%%%%%%%%%%%%%%%%%%%%%%%%%%%%%%%%%%%%%%
The energy levels of Sm-$4f$ states in SmFe$_{12}$
depend on ${\bm B}_{\rm ex}(T)$ and applied field ${\bm B}$.
Fig. \ref{fig:engalpha} shows the $B_{\rm ex}(T)/B_{\rm ex}(0)$
dependence of the energy levels for ${\bm e}^{TM}$=${\bm n}_c$,
which is a unit vector parallel to $c$-axis, and for ${\bm B}={\bm 0}$.
The data needed are given in TABLE \ref{table:CFP}.
As for SOI constant,
we use experimental value of $\lambda/k_B=\xi/5k_B=411$ K \cite{Elliott}.
At $B_{\rm ex}(T)=0$ the $S$ is strongly coupled with $L$
to form a Kramers doublet with a total angular momentum $J$
due to the large $LS$ coupling with fine CF splitting.
With increasing $B_{\rm ex}(T)/B_{\rm ex}(0)$,
the exchange field breaks the time-reversal symmetry and lift the degeneracy.

  \subsection{Phenomenological Model for $TM$ Sublattice}
  
For finite temperature magnetic properties of TM, we apply
a phenomenological formula
assuming uniform $M^{TM}(T)$ and $K_1^{TM}(T)$.
For $M^{TM}(T)$, we apply the Kuz'min formula \cite{Kuzminmag}:
\begin{align}
\frac{M^{TM}(T)}{M^{TM}(0)} &  =\frac{B_{\rm ex}(T)}{B_{\rm ex}%
(0)}=\alpha(T),\\
\alpha(T) &  =\left[  1-s\left(  \frac{T}{T_{\rm C}}\right)
^{3/2}-(1-s)\left(  \frac{T}{T_{\rm C}}\right)  ^{5/2}\right]
^{1/3},\label{eq:alpha}%
\end{align}
where, $T_{\rm C}$ is Curie temperature and $s$ is a fitting parameter.
The temperature dependence of $K_{1}^{TM}(T)$ has been expressed by an extended power
law \cite{Miura-pl}:
\begin{align}
\frac{K_{1}^{TM}(T)}{K_{1}^{TM}(0)}=&\alpha^{3}(T)+\frac{8}{7}C_{1}\left[
\alpha^{3}(T)-\alpha^{10}(T)\right]  \nonumber\\&+\frac{8}{7}C_{2}\left[  \alpha
  (T)^{3}-\frac{18}{11}\alpha(T)^{10}+\frac{7}{11}\alpha(T)^{21}\right],
\label{eq:KTMT}
\end{align}
where $C_{1}$ and $C_{2}$ are fitting parameters. 

In present study for SmFe$_{12}$ compound,
we use values of $s=0.01$ and $T_{\rm C}=555$ K in Eq. (\ref{eq:alpha}),
as used by Hirayama $et$ $al$. \cite{Hirayama2}. They showed that
the magnetization agrees well with experimental measurement for SmFe$_{12}$.
The values of $C_{1}$, $C_{2}$ and $V_0K_1^{TM}(0)$ in Eq. (\ref{eq:KTMT})
are determined
as $-0.263$, $-0.237$ and $47.7$ K, respectively,
by fitting the expression to observed data for YFe$_{11}$Ti
in Ref. \cite{Nikitin}.

\bigskip
\bigskip
\section{Method of Model Calculations}
\subsection{Statistical Method}

\label{sec:num} 

To calculate the finite temperature magnetic properties, we use the model
Hamiltonian and calculate MA and magnetic moment for Sm $4f$ electrons using
the statistical method for the partial system.
Using the eigenvalues of the Hamiltonian Eq. (\ref{eq:Heff}),
we express the free energy
density as,
\begin{align}
  G({\bm e}^{TM},T,{\bm B}) =&\frac{1}{V_0}\sum_{j=1}^{n_R}
  g_j({\bm e}^{TM},T,{\bm B})\nonumber\\
  &+K_{1}^{TM}(T)\sin^{2}\theta^{TM}-{\bm B}\cdot{\bm M}^{TM}(T),\label{eq:Gnum}\\
  g_j({\bm e}^{TM},T,{\bm B}) =&-k_{\rm B}T\ln Z_{j}({\bm e}^{TM},T,{\bm B}),\label{eq:gRnum}\\
Z_{j}({\bm e}^{TM},T,{\bm B}) =&\sum_{n}\exp\left[  -\frac{E_{n,j}%
({\bm e}^{TM},T,{\bm B})}{k_{\rm B}T}\right],
\end{align}
where $g_j({\bm e}^{TM},T,{\bm B})$ is Gibbs free energy for $R$-$4f$ partial system,
$E_{n,j}({\bm e}^{TM},T,{\bm B})$ and $Z_{j}({\bm e}^{TM},T,{\bm B})$
are the eigenvalue and the partition function of $j$-th $R$ Hamiltonian
$\hat{\cal H}_{R,j}$ [Eq. (\ref{eq:Heff})] for given
${\bm e}^{TM}$, respectively.
The direction of the $TM$ magnetization ${\bm e}^{TM}$ is
treated as an external parameter. The equilibrium condition of the system for
given $T$ and $\bm B$ is:
\begin{equation}
G({\bm e}_{0}^{TM},T,{\bm B}%
)=\min_{{\bm e}^{TM}}G({\bm e}^{TM},T,{\bm B}%
),\label{eq:Geq}%
\end{equation}
where ${\bm e}_{0}^{TM}$ is the direction of $TM$ sublattice magnetization
in the equilibrium. In practice, we determine the minimal $G({\bm e}^{TM}
,T,{\bm B})$ numerically by changing ${\bm e}^{TM}$.

The MA energy is given by the free energy $G({\bm e}^{TM},T,{\bm 0})$
with different directional vector ${\bm e}^{TM}$.
In the tetragonal symmetry, $g_j({\bm e}^{TM},T,{\bm 0})$ in
$G({\bm e}^{TM},T,{\bm 0})$ is
formally expressed as \cite{Kuzminlinear,Miura-pl}:
\begin{align}
g_j({\bm e}^{TM},T,{\bm 0})=&\sum_{p=1}^{\infty}\left[  k_{p,j}(T)+\sum
  _{q=1}^{\lfloor p/2\rfloor}k_{p,j}^{q}(T)\cos(4q\varphi^{TM})\right]\nonumber\\
  &\times\sin^{2p}\theta^{TM}+C(T),\label{eq:Gpheno}%
\end{align}
where $\theta^{TM}$ and $\varphi^{TM}$ are polar and
azimuthal angle of ${\bm e}^{TM}$, respectively, $\lfloor p/2\rfloor$
indicates the greatest integer of $p/2$, and
$k_{p,j}(T)$ and $k_{p,j}^{q}(T)$ are out-of-plane and in-plane MA constant for $j$-th $R$ ion.
The $C(T)$ is an angle independent constant.
The series expansion does not guarantee the convergence
\cite{Kuzmin2nd,Kuzminlimit}, however, for finite $p$, $k_{p,j}^{q}(T)$ can be
obtained from the comparison between Taylor series of $g_j({\bm e}^{TM},T,{\bm 0})$
of Eqs. (\ref{eq:gRnum}) and (\ref{eq:Gpheno}) with respect to $\theta
^{TM}$ for a fixed $\varphi^{TM}$ \cite{Sasaki,Miura-direct} as:%
\begin{align*}
g_j({\bm e}^{TM},T,{\bm 0}) 
=&g_j^{(0)}(T)+g_j^{(1)}(T)\theta^{TM}\\
&+\frac{1}{2!}g_j^{(2)}(T)(\theta^{TM})^{2}+\cdots,\\
g_j^{(n)}(T)=&\left.\frac{\partial^{n}g_j(\theta^{TM},\varphi^{TM},T,{\bm 0})}%
  {\partial(\theta^{TM})^{n}}\right|_{\substack{\theta^{TM}=0,\\\varphi^{TM}=\pi/8}},
\end{align*}
and
\begin{align*}
  g_j({\bm e}^{TM},T,{\bm 0}) =&k_{1,j}(T)(\theta^{TM})^{2}\\
  &+\left[  -\frac{2}{3!}%
k_{1,j}(T)+k_{2,j}(T)\right]  (\theta^{TM})^{4}+\cdots,
\end{align*}
respectively, which are resulting in
\begin{align}
k_{1,j}(T) &  =\frac{1}{2}g_j^{(2)}(T),\label{eq:K1num}\\
k_{2,j}(T) &  =\frac{1}{3}k_{1,j}(T)+\frac{1}{4!}g_j^{(4)}(T),\label{eq:K2num}%
\end{align}
etc.
  Using MA energy on the single $R$ ion in Eq. (\ref{eq:Gpheno}),
  the total MA constants are obtained as
\begin{align}
  K_1(T)&=\frac{1}{V_0}\sum_{j=1}^{n_R}k_{1,j}(T)+K_1^{TM}(T),&(p=1),
  \label{eq:K1tot}\\
  K^{(q)}_p(T)&=\frac{1}{V_0}\sum_{j=1}^{n_R}k_{p,j}^{(q)}(T),&(p\ge2),
  \label{eq:Kptot}
\end{align}
where $K_{p}(T)$ and $K_{p}^{q}(T)$ are out-of-plane and in-plane MA constant in whole system.

The orbital and spin components of the magnetic moment
of a single $R$ ion
in the equilibrium, can be calculated by:
\begin{align}
{\bm m}_{L,j}(T,{\bm B}) &  =-\mu_{\rm B}%
\sum_{n}\rho_{n,j}(T,{\bm B})\langle n,j|{\hat{\bm L}}%
|n,j\rangle, \label{eq:momnumL}\\%
{\bm m}_{S,j}(T,{\bm B}) &  =-2\mu_{\rm B}%
\sum_{n}\rho_{n,j}(T,{\bm B})\langle n,j|{\hat{\bm S}}%
|n,j\rangle, \label{eq:momnumS}%
\end{align}
respectively, where
$\rho_{n,j}(T,{\bm B})=\exp\left[-\beta E_{n,j}({\bm e}^{TM}_0,{\bm B})\right]
/Z_j({\bm e}^{TM}_0,T,{\bm B})$ and
$|n,j\rangle$ is the $n$-th eigenstate for $E_{n,j}({\bm e}_{0}^{TM},{\bm B})$, and
the total magnetization $M_{\rm s}(T,{\bm B})$ is given as,
\begin{align}
  {\bm M}_{\rm s}(T,{\bm B}) &  =\frac{1}{V_0}\sum_{j=1}^{n_R}
  {\bm m}_{j}(T,{\bm B})+M^{TM}(T){\bm e}_0^{TM},\label{eq:Magnum}
\end{align}
with ${\bm m}_j(T,{\bm B})={\bm m}_{L,j}(T,{\bm B})+{\bm m}_{S,j}(T,{\bm B})$.

Finally, to confirm the convergence of the probability weights
for excited-$J$ multiplet states at ${\bm B}={\bm 0}$,
we define a following weight function:
\begin{equation}
W_{J}(T)=\sum_{n,M}\rho_{n,j}(T,{\bm 0}) |\langle n,j|J,M \rangle|^2. \label{eq:WJ}
\end{equation}
In the case of SmFe$_{12}$ crystal, the value of $W_J(T)$ is independent of site index $j$.
%%%%%%%%%%%%%%%%%%%%%%%%%%%%%%%%%%%%%%%%%%%%%%%%%%%%%%
\begin{table}[ptb]
  \caption{Probability weight for each $J$-multiplet
    calculated by $W_J(T)$ in Eq. (\ref{eq:WJ}).
    For $J=13/2$ and $15/2$, $W_J(T)=0.0$.}
\begin{ruledtabular}
    \begin{tabular}{ccccc}
$J$& 5/2& 7/2& 9/2& 11/2
  \\ \hline
 $T$=0 & 0.93217 & 0.06548 & 0.00229 & 0.00005\\
 $T=T_{\rm C}$ & 0.90536 & 0.09049 & 0.00406 & 0.00009\\
\end{tabular}
\end{ruledtabular}
\label{table:WJ}%
\end{table}
%%%%%%%%%%%%%%%%%%%%%%%%%%%%%%%%%%%%%%%%%%%%%%%%%%%%%%%%%%%%
The results are shown in TABLE \ref{table:WJ}, which indicates good convergence
of weight for the number of the excited $J$-multiplets even at $T=T_{\rm C}=555$ K. 
Thus in the calculation using statistical method
for SmFe$_{12}$, we take the excited $J$-multiplets up to $J=9/2$. 
In the analytical calculation, the $J$-mixing effects are approximately
treated only for the lowest-$J$ multiplet by using unitary transformation.

\subsection{Analytical Method}

\label{sec:an} 

According to hierarchy of energy scale in $R$ intermetallic
compounds: $\hat{\cal H}_{\rm so}\gg\hat{\cal H}_{\rm ex}
\gg\hat{\cal H}_{\rm CF}\sim\hat{\cal H}_{\rm Z}$, we
develop an analytical method for finite temperature magnetic properties, which
enables us to connect the thermodynamic properties directly to our model
parameters based on electronic states. Practically, we generalize the
analytical expression of Gibbs free energy \cite{Yamada,Kuzminlinear}
  to include the effects of $J$-mixing using
  a first-order perturbation for the CF potential and Zeeman energy.
We also derive an analytical expression for the magnetization curve, which enables
us to estimate the CF potential using the observed results. The procedure of
the formalism consists of (i) construction of starting Hamiltonian for single
$R$ ion, (ii) approximation for diagonal matrix element of an effective
Hamiltonian, (iii) finite temperature perturbation for single $R$ ion, and (iv)
thermodynamic analysis. 

\subsubsection{Effective Lowest-$J$ Multiplet Hamiltonian \\for Single $R$ Ion}

To restrict $\hat{\cal H}_R$ in low-energy subspace for
$\hat{\cal H}_{\rm so}\gg\hat{\cal H}_{\rm ex}$, the
effective lowest-$J$ multiplet Hamiltonian $\hat{\cal H}_R^{{\rm eff}J}$
is obtained by unitary transformation and projection, where
the off-diagonal matrix elements between inter-$J$ multiplets become negligibly
small, and compensating term $\hat{\cal H}_{\rm mix}$ is added in
diagonal element for lowest-$J$ multiplet. We here introduce modified version
of effective Hamiltonian as explained below.

First, we define a rotational operator $\hat{\cal D}({\bm e}^{TM})$ which
transforms the quantization axis to ${\bm e}^{TM}$.
With this operator, the
Hamiltonian $\hat{\cal H}_R$ and $\hat{\cal H}_A$ ($A$=ex,
CF and Z) is transformed to:
\begin{align}
\hat{\cal D}^{\dagger}({\bm e}^{TM}&)\hat{\cal H}
_R\hat{\cal D}({\bm e}^{TM})\nonumber\\
&\equiv \hat{\cal H}_R'=\hat{\cal H}_{\rm so}'+\hat{\cal H}_{\rm ex}'%
+\hat{\cal H}_{\rm CF}'+\hat{\cal H}_{\rm Z}',
\end{align}
\begin{align}
\hat{\cal H}_{\rm so}'=& \frac{\lambda}{2}\left[  \hat{\bm J}%
^{2}-L(L+1)-S(S+1)\right]  ,\\
\hat{\cal H}_{\rm ex}'=& -2B_{\rm ex}(T)C_{0}^{(1)}(\hat{\bm S}),\label{eq:Hexp}\\
\hat{\cal H}_{\rm CF}'=&   \sum_{l,m,m'}B_{l}^{m}\Theta
_{l}^{L}[{\cal D}_{m,m'}^{(l)}({\bm e}^{TM})]^{\ast}C_{m'%
}^{(l)}(\hat{\bm L}),\\
\hat{\cal H}_{\rm Z}'=&  \mu_{\rm B} \sum_{m,m'}b_{-m}^{(1)}%
    [{\cal D}_{m,m'}^{(1)}({\bm e}^{TM})]^{\ast}
    \left[  C_{m'}^{(1)}(\hat{\bm L})+2C_{m'}^{(1)}(\hat{\bm S})\right],
\end{align}
where $\hat{\bm J}=\hat{\bm L}+\hat{\bm S}$,
$C_{q}^{(k)}(\hat{\bm A})$ is the spherical tensor operator with rank
$k$ for angular momentum $\hat{\bm A}$ \cite{Edomonds}, and $b_{m}%
^{(1)}$ is a magnetic field tensor: $b_{0}^{(1)}=B_{z}$ and $b_{\pm1}%
^{(1)}=-(\pm B_{x}+iB_{y})/\sqrt{2}$. ${\cal D}_{m,m'%
}^{(l)}({\bm e}^{TM})={\cal D}_{m,m'}^{(l)}(\varphi
^{TM},\theta^{TM},0)$ is the Wigner's $D$ function.
Now we apply a unitary transformation
(Schrieffer-Wolf transformation \cite{Schrieffer}) to $\hat{\cal H%
}_R'$,
\begin{equation}
e^{i\hat{\Omega}}\hat{\cal H}_R'e^{-i\hat{\Omega}}%
=\hat{\cal H}_R'+i\left[  \hat{\Omega},\hat{\cal H}%
_R'\right]  +O(\hat{\Omega}^{2}),\label{eq:SWtrans}%
\end{equation}
and introduce a projection operator $\hat{\cal P}_{J}=\sum_{M=-J}%
^{J}|J,M\rangle\langle J,M|$, by which the space of the $J$-multiplet is
restricted to
the lowest one. The operator $\hat{\Omega}$ is defined so as
to remove the first-order off-diagonal matrix elements for $J$ in
$\hat{\cal H}_R'$:

\begin{equation}
i\sum_{J'}\left[  \hat{\Omega},\hat{\cal P}_{J'}%
\hat{\cal H}_R'\hat{\cal P}_{J'}\right]
=\sum_{J'}\hat{\cal P}_{J'}\hat{\cal H}_R'\hat{\cal P}_{J'}-\hat{\cal H}_R'%
.\label{eq:Omegacomm}%
\end{equation}
Apparently, $\langle J,M|\hat{\Omega}|J,M'\rangle=0$.
The second term of the right-hand-side of Eq. (\ref{eq:SWtrans}) has now a
diagonal matrix with corrections to the diagonal elements in the original
$\hat{\cal H}_R'$. The second and higher-order terms in
$\hat{\Omega}$ are neglected. By inserting Eq. (\ref{eq:Omegacomm}) to Eq.
(\ref{eq:SWtrans}), we obtain
\begin{align}
\hat{\cal H}_R^{{\rm eff}J}= &  \hat{\cal P}_{J}e^{i\hat
{\Omega}}\hat{\cal H}_R'e^{-i\hat{\Omega}}\hat{\cal P}%
_{J}\equiv\hat{\cal H}_R^{J}+\hat{\cal H}_{\rm mix},\label{eq:HR}\\
\hat{\cal H}_R^{J}= &  \hat{\cal P}_{J}\hat{\cal H}%
_R'\hat{\cal P}_{J}=E_{J}+\hat{\cal H}_{\rm ex}^{J}%
+\hat{\cal H}_{\rm CF}^{J}+\hat{\cal H}_{\rm Z}^{J},\\
\hat{\cal H}_{\rm mix}= &  \frac{i}{2}\hat{\cal P}_{J}\left[
  \hat{\Omega},\hat{\cal H}_R'\right]  \hat{\cal P}_{J},\label{eq:Hmix}%
\end{align}
where $E_J=\lambda[J(J+1)-L(L+1)-S(S+1)]/2$ and 
$\hat{\cal H}_A^{J}=\hat{\cal P}_{J}%
\hat{\cal H}_A'\hat{\cal P}_{J}$ ($A$=ex, CF and Z).
We here classify analytical models depending on the
  approximation to the matrix element of $\hat{\Omega}$ for $J\ne J'$
  in Eq. (\ref{eq:Omegacomm}) as follows:
\begin{itemize}
\item model A: Lowest-$J$ multiplet without mixing as:\\
  \begin{equation*}
    \langle J,M|\hat{\Omega}^{\rm A}|J',M'\rangle=0,
  \end{equation*}
\item model B: Effective lowest-$J$ multiplet with mixing as:\\
  \begin{equation*}
  \langle J,M|\hat{\Omega}^{\rm B}|J',M'\rangle
    =i\frac{\langle J,M|\hat{\cal H}_1|J',M'\rangle}{E_{J'}-E_J}\nonumber,
    \end{equation*}
\item model C: Modified effective lowest-$J$ multiplet with mixing (present study) as:.
\begin{align*}
  \langle J,M|&\hat{\Omega}^{\rm C}|J',M'\rangle\\
  =&i\frac{\langle J,M|\hat{\cal H}_1|J',M'\rangle}{E_{J'}-E_J}-\frac{i}{(E_{J'}-E_J)^2}\\
  &\times\sum_{M''}\left[\langle J,M|\hat{\cal H}_1|J',M''\rangle\langle J',M''|\hat{\cal H}_1|J',M'\rangle\right.\\
   &\left.\qquad\quad -\langle J,M|\hat{\cal H}_1|J,M''\rangle\langle J,M''|\hat{\cal H}_1|J',M'\rangle\right],
  \end{align*}
\end{itemize}
where $\hat{\cal H}_1\equiv \hat{\cal H}_R'-\hat{\cal H}_{\rm so}'$.
The approximations are referred to as model A, B and C, hereafter.
By using $\hat{\Omega}^{\rm B}$, Magnani $et$ $al$. derived the effective lowest-$J$ multiplet Hamiltonian
\cite{Magnani} and Kuz'min had also derived an equivalent approximation for
anisotropy constants \cite{Kuzminmix}. In the latter work, it was pointed out
that the approximations of the models A and B are not applicable to the Sm compounds
due to relatively small $\lambda$.
In the present study, we have modified $\hat{\Omega}^{\rm B}$ to $\hat{\Omega}^{\rm C}$.

\subsubsection{Approximation for Diagonal Matrix Element of $\hat{\cal H%
}_R^{{\rm eff}J}$} \label{sec:diag}

The energy levels for $4f$ electron system are obtained by the exact
diagonalization of $\hat{\cal H}_R$ in Eq. (\ref{eq:Heff}),
and the diagonal matrix elements of
$\hat{\cal H}_R^{{\rm eff}J}$ can be expressed as:
\begin{align}
  \langle J,M|&\hat{\cal H}_R^{{\rm eff}J}|J,M\rangle\nonumber\\
  &=  \langle J,M|\hat
{\cal H}_R'|J,M\rangle+\langle J,M|\hat{\cal H}_{\rm mix}%
|J,M\label{eq:HReffdiag}\rangle,
\end{align}
through two unitary transformations by $\hat{\cal D}
({\bm e}^{TM})$ and $e^{-\hat{\Omega}}$.
The first term in Eq. (\ref{eq:HReffdiag}) can be obtained by using
the relation ${\cal D}_{m,0}^{(l)}(\varphi^{TM},\theta^{TM},0)=Y_{l}^{m}%
(\theta^{TM},\varphi^{TM})$ and Wigner Eckert theorem \cite{Edomonds},
\begin{align}
  \langle J,M|&\hat{\cal H}_R'|J,M\rangle\nonumber\\  = &  E_{J}-2(g_{J}%
-1)\mu_{\rm B}B_{\rm ex}(T)\langle J,M|C_{0}^{(1)}(\hat{\bm J})|J,M\rangle
\nonumber\\
&  +\sum_{l,m}A_{l}^{m}\langle r^{l}\rangle\Theta_{l}^{J}\frac{t_{l}^{m}({\bm
    e}^{TM})}{a_{l,m}}\langle J,M|C_{0}^{(l)}(\hat{\bm J})|J,M\rangle\nonumber\\
&+\mu_{\rm B}g_J\left({\bm e}^{TM}\cdot {\bm B}\right)\langle J,M|C_{0}^{(1)}(\hat{\bm J})|J,M\rangle,
\end{align}
where $\Theta_{l}^{J}$ is the Stevens factor \cite{Stevens, Hutchings}.
By using the model C with $\hat{\Omega}^{\rm C}$,
the second term in Eq. (\ref{eq:HReffdiag}) is approximated as,
\begin{align}
  \langle J,&M|\hat{\cal H}_{\rm mix}|J,M\rangle\nonumber\\
  \sim&-\frac{1}{\Delta_{\rm so}}\langle
J,M|\hat{\cal H}_{\rm ex}'|J+1,M\rangle\nonumber\\
&\times\langle J+1,M|\hat{\cal H}_{\rm ex}'+2\hat{\cal H}%
_{\rm CF}'+2\hat{\cal H}_{\rm Z}'|J,M\rangle
\nonumber\\
&  \times\left[  1-\frac{\langle J+1,M|\hat{H}_{\rm ex}'|J+1,M\rangle
    -\langle J,M|\hat{H}_{\rm ex}'|J,M\rangle}{\Delta_{\rm so}}\right],
\label{eq:HmixC}%
\end{align}
where $\Delta_{\rm so}=\lambda(J+1)$. 
Contributions from $\hat{\cal H}_{\rm CF}'$ and
$\hat{\cal H}_{\rm Z}'$ are neglected in the second term of
the square bracket. By using Wigner-Eckert theorem \cite{Edomonds} and the relation for
products of the matrix elements of the spherical tensor operators given by Eq.
(5) in chapter 12. of Ref. \cite{Varshalovich}, the diagonal matrix element is
expressed as follows:
\begin{align}
  \langle J,M&|\hat{\cal H}_{\rm mix}|J,M\rangle\nonumber\\
  =&  -\Delta_{\rm ex}(T)\frac{L+1}{3S}\langle J,M|{\cal T}_{1}(\hat{\bm J}%
)|J,M\rangle\nonumber\\
&  -\sum_{l,m}A_{l}^{m}\langle r^{l}\rangle\Xi_{l}^{J}\frac{t_{l}^{m}%
({\bm e}^{TM})}{a_{l,m}}\frac{l(l+1)}{2l+1}\langle J,M|{\cal T}%
_{l}(\hat{\bm J})|J,M\rangle\nonumber\\
&  +\left(  {\bm e}^{TM}\cdot{\bm B}\right)  \frac{2(L+1)}{3(J+1)}\langle
J,M|{\cal T}_{1}(\hat{\bm J})|J,M\rangle, \label{eq:HmixX}%
\end{align}
where $\Delta_{\rm ex}(T)=-2(g_{J}-1)\mu_{\rm B}B_{\rm ex}(T)$.
We here use the relation $J=L-S$ assuming $R$ as light rare-earth and
$\Xi_{6}^{J}=-2^{2}/(3^{3}\times7\times11)$ and $-2^{2}\times
17/(3^{5}\times7\times11^{2})$ for Ce$^{3+}$ and Sm$^{3+}$, respectively, and
$\Xi_{l}^{J}=\Theta_{l}^{J}$ in the other cases.

More explicit expression of ${\cal T}_{1}(\hat{\bm J})$ depends on
further approximations. So far two approximations have been adopted; one
completely neglect the term $\langle J,M|\hat{\cal H}_{\rm mix}|J,M\rangle$,
that is, $\hat{\cal H}_{\rm mix}=0$
\cite{Kuzminlinear}, and the other is an approximation to neglect the second
term \ in the square bracket in Eq. (\ref{eq:HmixC}) which was adopted by
Kuz'min \cite{Kuzminmix} and Magnani $et$ $al$ \cite{Magnani}.
According to the model approximations of $\hat{\Omega}^X$ with $X=$A, B, and C,
the quantities ${\cal T}%
_{l}(\hat{\bm J})$ are denoted as
${\cal T}_{l}^X(\hat{\bm J})$ with $X$=A, B, and C.
Clearly $\hat{\cal T}_{l}^{\rm A}=0$, and for $X$=B and C,%
\begin{align}
  {\cal T}_{l}^{\rm B(C)}(\hat{\bm J})=&\frac{\Delta_{\rm ex}(T)}{\Delta_{\rm so}}
  \left[  \frac{2J+l+1}{2}{\cal V}_{l-1}%
    ^{\rm B(C)}(\hat{\bm J})\right.\nonumber\\
   &\left. -\frac{2}{2J+l+2}{\cal V}_{l+1}^{\rm B(C)}
            (\hat{\bm J})\right],\label{A:Tl}
\end{align}
with
\begin{align}
{\cal V}_{l}^{\rm B}(\hat{\bm J}) =&C_{0}^{(l)}(\hat{\bm J}%
),\nonumber\\
{\cal V}_{l}^{\rm C}(\hat{\bm J}) =&C_{0}^{(l)}(\hat{\bm J}%
)+\frac{\Delta_{\rm ex}(T)}{\Delta_{\rm so}}\frac{L+S+1}%
{S(J+2)}\nonumber\\
&  \times\left[\frac{l(2J-l+1)(2J+l+1)}{4(2l+1)}C_{0}^{(l-1)}(\hat{\bm J})\right.\nonumber\\
&\qquad\left.+\frac{l+1}{2l+1}C_{0}^{(l+1)}(\hat{\bm J})\right]  , \label{A:V1}%
\end{align}
where we formally set $C_{0}^{(-1)}(\hat{\bm J})=0$.

The energy levels $E_{n}$ for $4f$ electron system, which consist of the
lowest energy $E_{1}$ to the $2J$-th excited energy $E_{2J+1}$, are now
expressed as,
\begin{equation}
E_{M}^X=\langle J,M|\hat{\cal H}_R'|J,M\rangle+\langle
J,M|\hat{\cal H}_{\rm mix}^X|J,M\rangle,\label{eq:EMX}%
\end{equation}
($X$=A, B, and C) with $M=-J$ to $J$ for the model A, B, and C.

%%%%%%%%%%%%%%%%%%%%%%%%%%%%%%%%%%%%%%%%%%%%%%%%%%%%%%%%%%%%%%%%%%%%%%%%%%%%%%
\begin{figure}[htbp]
\begin{center}
\includegraphics[width=8.0cm]{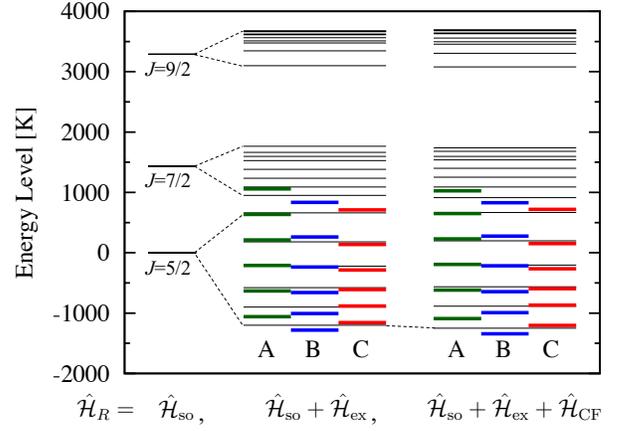}
\end{center}
\par
\caption{(color online) Calculated energy levels of the Sm-$4f$
  states in SmFe$_{12}$ at ${\bm B}={\bm 0}$.
  Analytical results $E^{X}_M$ with $X$=A, B, and C for corresponding model approximations
  in Eq. (\ref{eq:EMX})
  are given by thick green (A), blue (B)
  and red lines (C), respectively.
  To clarify the contributions from $\hat{\cal
{\cal H}}_{\rm so},\hat{\cal H}_{\rm ex},$ and $\hat{\cal H%
}_{\rm CF}$, we take original Hamiltonian $\hat{\cal H}_R$ as
$\hat{\cal H}_R=\hat{\cal H}_{\rm so}$, $\hat{\cal H%
}_{\rm so}+\hat{\cal H}_{\rm ex}$, and $\hat{\cal H%
}_{\rm so}+\hat{\cal H}_{\rm ex}+\hat{\cal H}%
_{\rm CF}$. The numerically exact results are also shown by thin black
lines.}%}%
\label{fig:englevel}%
\end{figure}
%%%%%%%%%%%%%%%%%%%%%%%%%%%%%%%%%%%%%%%%%%%%%%%%%%%%%%%%%%%%%%%%%%%%%%%%%%%%%%

Fig. \ref{fig:englevel} shows the diagonal matrix element $E_{M}%
^X$ ($X$=A, B, and C) of the effective lowest-$J$ multiplet
Hamiltonian $\hat{\cal H}_R^{{\rm eff}J}$ at $T=0$ in Eq. (\ref{eq:EMX}).
Note that CF coefficients and exchange fields are
determined by the first principles, and the same values are used for models A,
B and C. The results are compared with
the exact results. To distinguish the contribution from each
$\hat{\cal H}_{\rm so},$ $\hat{\cal H}_{\rm ex}$, and
$\hat{\cal H}_{\rm CF}$ in $\hat{\cal H}_R$ of Eq.
(\ref{eq:Heff}), the original Hamiltonian $\hat{\cal H}_R$ is
taken as $\hat{\cal H}_{\rm so}$, $\hat{\cal H}_{\rm so}+
\hat{\cal H}_{\rm ex}$, or $\hat{\cal H}_{\rm so}%
+\hat{\cal H}_{\rm ex}+\hat{\cal H}_{\rm CF}$.

Let us first describe the characteristics for the result $\hat{\cal H%
}_R =\hat{\cal H}_{\rm so}+\hat{\cal H}%
_{\rm ex}$. In model A, the sixfold degeneracy of energy levels given by
$\hat{\cal H}_{\rm so}$ splits into equi-energy levels as
$E_{M}^{\rm A}=E_{J}+\Delta_{\rm ex}(0)M$. In model B, the
equi-energy levels shift to lower energy states by $J$-mixing term,
$E_{M}^{\rm B}=E_{J}+\Delta_{\rm ex}(0)M-|\langle J,M|\hat
{\cal H}_{\rm ex}|J+1,M\rangle^{2}/\Delta_{\rm so}.$ In model
C, the energy shifts, which was over-estimate
by the $J$-mixing term, are corrected.

The results obtained by $\hat{\cal H}_R=\hat
{\cal H}_{\rm so}+\hat{\cal H}_{\rm ex}+\hat{\cal H}_{\rm CF}$
show that the effect of CF potentials on the energy levels is
weak, as expected, and they reproduce the results obtained by
the numerical exact diagonalization method as shown in Fig. \ref{fig:englevel}.

\subsubsection{Finite Temperature Perturbation for Single $R$ Ion}
\label{sec:1PT}

We apply the first-order perturbation at finite temperature assuming
$\hat{\cal H}_{\rm ex}^{J}\gg\hat{\cal H}_{\rm CF}%
^{J}+\hat{\cal H}_{\rm Z}^{J}+\hat{\cal H}_{\rm mix}^X$.
The unperturbed and perturbed Hamiltonians are $\hat
{\cal H}_{\rm ex}^{J}=\Delta_{\rm ex}(T)C_{0}^{(1)}%
(\hat{\bm J})\equiv\hat{\cal H}^{(0)}$ and $\hat{\cal H%
}_{\rm CF}^{J}+\hat{\cal H}_{\rm Z}^{J}+\hat{\cal H%
}_{\rm mix}\equiv\hat{\cal H}'$, respectively.
Note that $\hat{\cal H}_{\rm so}$ is effectively taken into account in
the $J$-multiplet formation of the $R$ ion. The approximated
Gibbs free energy for $R$-$4f$ partial system on $j$-th $R$ site
up to first-order perturbation is formally expressed as
$g_j({\bm e}^{TM},T,{\bm B})=-k_{\rm B}\ln Z_0(T)+\sum_{M}\rho_{M}^{(0)}(T)\langle J,M|
\hat{\cal H}'|J,M\rangle$,
where $E_M^{(0)}(T)=\Delta_{\rm ex}(T)M$,
$Z_0(T)=\sum_{M}\exp[-\beta E_{M}^{(0)}(T)]$,
and $\rho_{M}^{(0)}(T)=\exp[-\beta E_{M}^{(0)}(T)]/Z_0(T)$. 
More explicitly, it is given as,
\begin{align}
g({\bm e}^{TM},T,{\bm B})=&k_{\rm B}T\sum_{M}\rho_{M}^{(0)}(T)\ln
\rho_{M}^{(0)}(T)\nonumber\\
&+\sum_{M}\rho_{M}^{(0)}(T)E_{M},
\end{align}
by using $E_{M}$ in Eq. (\ref{eq:EMX}). It is noted that $g({\bm e}%
^{TM},T,{\bm B})$ is model dependent because $E_M$ equals to
$E_{M}^{\rm A}$, $E_{M}^{\rm B}$ or $E_{M}^{\rm C}$,
corresponding to the model adopted.

By using Helmholtz free energy $f({\bm e}^{TM},T)$ for $R$-$4f$
partial system, the Gibbs free energy in the modified effective lowest-$J$
model is given as,
\begin{align}
g({\bm e}^{TM},T,{\bm B})= &  f({\bm e}%
^{TM},T)-m(T){\bm e}^{TM}\cdot{\bm B},\label{eq:GRan}\\
m(T)  =&\mu_{\rm B}\left[g_{J}JB_{J}^{1}(x)-\frac{2(L+1)}{3(J+1)}T_J^{1}(x)\right],\label{eq:moman}
\end{align}
with
\begin{align}
f({\bm e}^{TM},T)=&k_{\rm B}T\sum_{M}\rho_{M}^{(0)}(T)%
\ln\rho_{M}^{(0)}(T)\nonumber\\
&+f_{\rm ex}(T)+f_{\rm CF}({\bm e}^{TM},T),\label{eq:FRan}\\
f_{\rm ex}(T)=&-\Delta_{\rm ex}(T)\left[  JB_{J}^{1}(x)+\frac{L+1}{3S}T_J^{1}(x)\right],
\label{eq:FRexan}\\
f_{\rm CF}({\bm e}^{TM},T)=&\sum_{l,m}A_{l}^{m}\langle r^{l}\rangle\Xi_{l}^{J}\frac{t_{l}%
  ^{m}({\bm e}^{TM})}{a_{l,m}}\nonumber\\
&\times\left[  J^{l}B_{J}^{l}(x)+
  \frac{l(l+1)}{2l+1}T_J^{l}(x)\right].\label{eq:FRCFan}
\end{align}
Here $x\equiv J\Delta_{\rm ex}(T)/k_{\rm B}T$, and the model
dependence appears in $T_{J}^{l}(x)$, which is denoted as $T_{J}%
^{l,X}(x)$ with $X$=A, B or C. For $X$=A, $T_{J}^{l,\rm{A}}(x)=0$ and
for $X$=B and C,
\begin{align}
T_J^{l,{\rm B(C)}}(x)=&\frac{\Delta_{\rm ex}(T)}{\Delta_{\rm so}}
\left[  \frac{2J+l+1}{2}V_J^{l-1,{\rm B(C)}}(x)\right.\nonumber\\
  &\qquad\qquad\left.-\frac{2}{2J+l+2}V_J^{l+1,{\rm B(C)}}(x)\right]  \label{eq:Tlm},
\end{align}
with
\begin{align}
V_J^{l,{\rm B}}(x)=& J^{l}B_{J}^{l}(x),\\
V_J^{l,{\rm C}}(x)=& J^{l}B_{J}^{l}(x)-\frac{\Delta_{\rm ex}(T)}
{\Delta_{\rm so}}\frac{L+S+1}{S(J+2)}\nonumber\\%
&\times\left[  \frac{l(2J-l+1)(2J+l+1)}{4(2l+1)}J^{l-1}B_{J}^{l-1}(x)\right.\nonumber\\
  &\qquad\left.+\frac{l+1}{2l+1}J^{l+1}B_{J}^{l+1}(x)\right]  ,\label{eq:VlmC}%
\end{align}
where $B_{J}^{l}(x)$ is the generalized Brillouin function
\cite{Kuzminlinear} defined by $(-1)^{l}J^{l}B_{J}^{l}(x)=\langle C_{0}%
^{(l)}(\hat{\bm J})\rangle_0$ with $x=J\Delta_{\rm ex}(T)/k_{\rm B}T$ for $l\geq0$, where $\langle
\hat{\bm A}\rangle_0=\sum_{M}\rho_{M}^{(0)}(T)\langle J,M|\hat{\bm A}|J,M\rangle$.
The analytical expression of $B_J^{l}(x)$ is given in Ref. \cite{Magnani}
and $T_J^{l,\rm A}(x)=0$, 
$T_J^{l,\rm B}(x)$ and $T_J^{l,\rm C}(x)$ are linear combination of $B_J^{l-1}(x)$ and $B_J^{l+1}(x)$,
and $B_J^{l-2}(x)$, $B_J^{l}(x)$, and $B_J^{l+2}(x)$, respectively,
as shown in Eq. (\ref{eq:Tlm}).

Because of the first-order perturbation for $\hat{\cal H}_{\rm Z}'$,
an analytical expression of the magnetic moment $m(T)$ is
obtained as $m(T)=m_{L}(T)+m_{S}(T)$ with:
\begin{align}
  m_{L}(T) &  =\mu_{\rm B}\left[  \frac{L+1}{J+1}JB_{J}^{1}(x)+
    \frac{2(L+1)}{3(J+1)}T_J%
^{1}(x)\right]  ,\label{eq:momLan}\\
  m_{S}(T) &  =-2\mu_{\rm B}\left[  \frac{S}{J+1}JB_{J}^{1}(x)+
   \frac{2(L+1)}{3(J+1)}T_J%
^{1}(x)\right]  ,\label{eq:momSan}%
\end{align}
where $m_{L}(T)$ and $m_{S}(T)$ are orbital and spin component of magnetic
moment on the $R$ ion. It is noted that $m_{L}(T)$ and $m_{S}(T)$ are model
dependent because of the model dependence of $T_{J}^{l}(x)$ as shown above.

Within the finite temperature perturbation theory, the angular ${\bm e}^{TM}$
dependent part of single $R$ ion free energy $f({\bm e}^{TM},T)$
in Eq. (\ref{eq:FRan}) with the tetragonal symmetry
can be written by:
\begin{align}
  f({\bm e}^{TM},T)=&k_1(T)\sin^2\theta\nonumber\\
  &+\left[k_2(T)+k_{2}^{1}(T)\cos4\varphi^{TM}\right]
  \sin^{4}\theta^{TM}\nonumber\\
  &+\left[k_3(T)+k_{3}^{1}(T)\cos4\varphi^{TM}\right]
  \sin^{6}\theta^{TM}\nonumber\\
  &+C(T),
\label{eq:FRCFan2}%
\end{align}
which is a truncated form of $g({\bf e}^{TM},T,{\bm 0})$ in Eq. (\ref{eq:Gpheno}).
The $C(T)$ is an angle independent constant. 
For example, the leading anisotropy constants for a
trivalent magnetic light $R$ ion (Ce$^{3+}$, Pr$^{3+}$, Nd$^{3+}$, Pm$^{3+}$,
and Sm$^{3+}$) can be written as follows:
\begin{align}
k_{1}(T) &  =-3\left[  J^{2}B_{J}^{2}(x)+\frac{6}{5}%
T_J^{2}(x)\right]  A_{2}^{0}\langle
r^{2}\rangle\Xi_{2}^{J}\nonumber\\
&  -40\left[  J^{4}B_{J}^{4}(x)+\frac{20}{9}%
T_J^{4}(x)\right]  A_{4}^{0}\langle
r^{4}\rangle\Xi_{4}^{J}\nonumber\\
&  -168\left[  J^{6}B_{J}^{6}(x)+\frac{42}{13}%
T_J^{6}(x)\right]  A_{6}^{0}\langle
r^{6}\rangle\Xi_{6}^{J},\label{eq:K1an}\\
k_{2}(T) &  =35\left[  J^{4}B_{J}^{4}(x)+\frac{20}{9}
T_J^{4}(x)\right]  A_{4}%
^{0}\langle r^{4}\rangle\Xi_{4}^{J}\nonumber\\
&  +378\left[  J^{6}B_{J}^{6}(x)+\frac{42}{13}%
T_J^{6}(x)\right]  A_{6}^{0}\langle
r^{6}\rangle\Xi_{6}^{J}.\label{eq:K2an}%
\end{align}
All terms of MA constants $k^{(q)}_{p}(T)$
in model A,B, and C are given
by linear terms with respect to $A_{l}^{m}\langle r^{l}\rangle$.

\begin{figure}[htbp]
\begin{center}
\includegraphics[width=8cm]{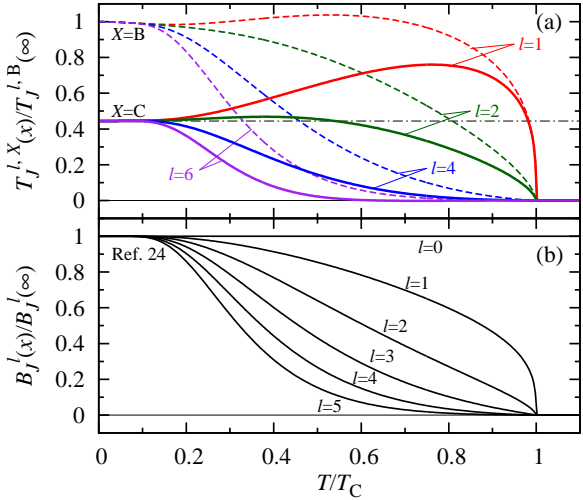} 
\end{center}
\caption{(color online) Temperature dependence of
  (a) $T_J^{l, \rm B(C)}(x)$ scaled by $T_J^{l, {\rm B}}(\infty)>0$
  in Eq. (\ref{eq:Tlm}) for model B(C) with broken(solid) curves and
  (b) generalized Brillouin function $B_J^l(x)/B_J^l(\infty)$ \cite{Kuzminlinear}
  with $J=5/2$ and $x=J\Delta_{\rm ex}(T)/k_{\rm B}$, where the
    temperature is scaled by Curie temperature $T_{\rm C}$.
  The dashed-dotted line represent the value of $R_J$ (see text).
}%
\label{fig:Tlm_scaled}%
\end{figure}
We may rewrite the approximations used and adopted in the present
formalism by using  $T_J^{l,X}(x)$ in Eq. (\ref{eq:Tlm}) as follows: 

\begin{itemize}
\item model A: Lowest-$J$ multiplet without mixing as $\Delta_{\rm ex}(T)/\Delta_{\rm so}=0$
or $T_J^{l, {\rm A}}(x)=0$ 
\cite{Kuzminlinear} 
\item model B: Effective lowest-$J$ multiplet with mixing as
$\left[\Delta_{\rm ex}(T)/\Delta_{\rm so}\right]^2=0$ or $T_J^{l, {\rm B}}(x)$ \cite{Kuzminmix,Magnani} 
\item model C: Modified effective lowest-$J$ multiplet with mixing as $T_J^{l,{\rm C}}(x)$ (present study). 
\end{itemize}

At $T=0$, we have found a following simple relation holds
between $T_J^{l, {\rm C}}(\infty)$ and $T_J^{l, {\rm B}}(\infty)$
as:
\begin{align}
  R_J=\displaystyle \frac{T_J^{l, {\rm C}}(\infty)}{T_J^{l, {\rm B}}(\infty)}=
  1-\frac{\Delta_{\rm ex}(0)}{\Delta_{\rm so}}\frac{(L+S+1)J}{S(J+2)}.
\label{eq:RJ}
\end{align}
Because $R_J$ is independent of $l$, relations among the models $X$=A, B, and C on
$m_{L,S}^X(0)$ and $k_{p}^{(q,)X}(0)$
can be generally expressed as follows:
\begin{align}
  m_{L,S}^{\rm C}(0)&=m_{L,S}^{\rm A}(0)+R_J\left[m_{L,S}^{\rm B}(0)-m_{L,S}^{\rm A}(0)\right],\label{eq:momRJ}\\
  k_{p}^{(q,)\rm C}(0)&=k_p^{(q,)\rm A}(0)+R_J\left[k_p^{(q,)\rm B}(0)-k_p^{(q,)\rm A}(0)\right].\label{eq:kRJ}
\end{align}

  At finite temperatures, $T_J^{l,{\rm B(C)}}(x)$ for $J=5/2$
  scaled by $T_J^{l,{\rm B}}(\infty)>0$
  are shown in Fig. \ref{fig:Tlm_scaled}(a) for the SmFe$_{12}$ compound.
Here, $\Delta_{\rm ex}(T)/\Delta_{\rm so}$ is taken to be $0.206\alpha(T)$.
  For comparison purpose, we also show the $B_J^{l}(x)/B_J^l(\infty)$ in Fig. \ref{fig:Tlm_scaled}(b).
$B_J^{l}(x)$ decays faster than $T_J^{l,{\rm B(C)}}(x)$ with increasing temperature.  
Thus the $J$-mixing effects included in $T_J^{l,{\rm B(C)}}(x)$ remain
even at high temperatures.

\subsubsection{Thermodynamic Analysis}

Finally we investigate the thermodynamical instability by using the
thermodynamic relation between Gibbs and Helmholtz free energy, which
explicitly contains the CF potentials and the exchange field determined by first
principles. We have to note that above the room temperature the exchange
contribution $\hat{\cal H}_{\rm ex}$ decreases with increasing
temperature at a rate proportional to $\alpha(T)$, so the energy hierarchy
is changed and thermal fluctuation effects have to be considered as
$k_{\rm B}T\gg\hat{\cal H}_{\rm CF}\sim\hat{\cal H%
}_{\rm ex}$. Even in this case, the formulation derived here based on
generalized Brillouin function holds as shown by Kuz'min in Refs.
\cite{Kuzminmix,Kuzminlimit}. In this thermodynamic analysis,
we use the model C. 

By applying the finite temperature perturbation theory to the lowest-$J$
multiplet Hamiltonian, the approximated Gibbs free energy density for whole
system can be expressed as:
\begin{align}
G({\bm e}^{TM},T,{\bm B}) &  =F({\bm e}^{TM}%
,T)-{\bm M}_{\rm s}(T)\cdot{\bm B},\label{eq:Gan}\\
F({\bm e}^{TM},T) &  =\frac{1}{V_0}\sum_{j=1}^{n_R}f_{j}({\bm e}^{TM},T)+K_{1}%
^{TM}(T)\sin^{2}\theta^{TM},\label{eq:Fan}\\
{\bm M}_{\rm s}(T) &  =\left[\frac{1}{V_0}\sum_{j=1}^{n_R}{m}_j(T)+M^{TM}(T)\right]
{\bm e}^{TM},\label{eq:Magan}%
\end{align}
where $F({\bm e}^{TM},T)$ is Helmholtz free energy density for whole
system with model C and $f_j({\bm e}^{TM},T)$ and $m_{j}(T){\bm e}^{TM}$
are corresponding energy for $4f$-shell and expectation value of magnetic moment on $j$-th $R$ ion given
in Eq. (\ref{eq:FRan}) and Eq. (\ref{eq:moman}), respectively. The
temperature dependence of $G({\bm e}^{TM},T,{\bm B})$ can be expressed as the
linear combination of the generalized Brillouin functions for $R$ ion
$B_{J}^{l}(J\Delta_{\rm ex}/k_{\rm B}T)$ and the temperature coefficient
for $TM$ ion $\alpha(T)$ in Eq. (\ref{eq:alpha}). The equilibrium condition is
the same as Eq. (\ref{eq:Geq}), where ${\bm e}_{0}^{TM}$ becomes the
direction of total magnetization in the equilibrium. We can also analyze the
instability of magnetic metastable states, which are crucially important in
permanent magnetic materials. The metastable condition is $\delta
G(T,{\bm e}^{TM},{\bm B})>0$ for given $T$ and ${\bm B}$ with $|e^{TM}|=1$.

The MA constants in whole system are obtained by combining the contribution from $R$ sublattice
in Eq. (\ref{eq:FRCFan2}) with Fe sublattice
same as Eqs. (\ref{eq:K1tot}) and (\ref{eq:Kptot}).
$K_{1}(T)$ can be substituted into the so called
Kr\"{o}nmuller equation \cite{Kronmuller1,Kronmuller-text} to obtain the
coercive field
\begin{align}
B_{\rm c}(T) &  =\alpha B_{\rm N}(T)-N_{\rm eff}M_{\rm s}(T),\\
B_{\rm N}(T) &  =\frac{2K_1(T)}{M_{\rm s}(T)},\label{eq:BN}%
\end{align}
where $B_{\rm c}(T)$ and $B_{\rm N}(T)$ are coercive and nucleation field, respectively.
$\alpha(<1)$ is microstructural parameter and $N_{\rm eff}$ is local effective
demagnetization factor \cite{Kronmuller-text}. The $B_{\rm N}(T)$ gives
upper limit of $B_{\rm c}(T)$.

\section{Calculated Results for SmFe$_{12}$}

\subsection{Valence Mechanism of Magnetic Anisotropy}

We first calculate the charge density distribution and Coulomb potential at 0
K on constituent atoms of SmFe$_{12}$ lattice (Fig. \ref{fig:struct}) using the first principles. The
calculated results determine the values of CF acting on $4f$ electrons, the
magnitude of the exchange field $B_{\rm ex}(0)$ acting on the $J$, and the
magnitude of $TM$ sublattice magnetization. These values are used for parameter
values in the model Hamiltonian. The contribution to the CF from the charge
density distribution inside (outside) the muffin-tine sphere radius is called
"valence (lattice) contribution" \cite{Hummler0}. If the CF is dominated by
the former contribution, we call the mechanism of the MA "valence mechanism"
\cite{Tsuchiura1}. 

\begin{figure*}[htbp]
\begin{center}
\includegraphics[width=15.0cm]{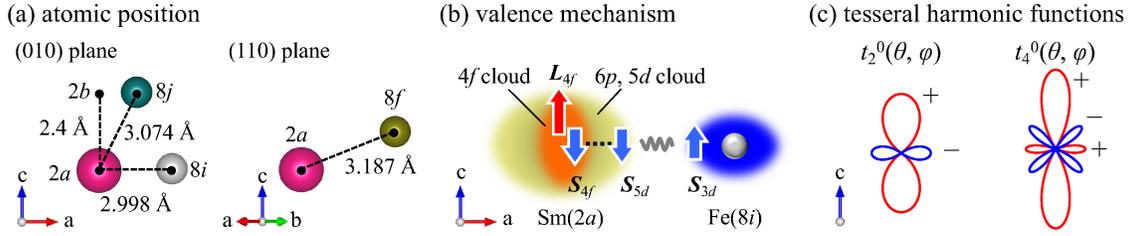}
\end{center}
\caption{(color online) (a) Atomic position of first ($8i$), second ($8j$), and third ($8f$)
  neighbor Fe atoms of Sm ion in SmFe$_{12}$, (b)
illustration of valence mechanism \cite{Tsuchiura1} in SmFe$_{12}$, and (c)
typical tesseral harmonic functions as basis of CF Hamiltonian, where signs
represent the phase.}%
\label{fig:pos_dis_tlm}%
\end{figure*}

The charge density distributions of single $R$ ion are approximately replaced
with charge density on atomic orbitals of $6p$ and $5d$ states. To evaluated the
valence contribution to CF parameters $A_{l}^{0}\langle r^{l}\rangle(val)$, we
introduce distribution parameters $\Delta n_{6p}^{(2)},\Delta n_{5d}^{(2)}$
\cite{Coehoorn,Sakuma} and $\Delta n_{5d}^{(4)}$ defined as,
\begin{equation}
\Delta n_{n'l'}^{(l)}=\frac{4\pi}{2l+1}a_{l,0}\sum_{m'}\int
d\Omega\ t_{l}^{0}(\theta,\varphi)|t_{l'}^{m'}(\theta
,\varphi)|^{2}n_{n'l',m'},\label{eq:dntlm}%
\end{equation}
where $\Omega$ is the solid angle and $m'$ indicates the multiple
orbitals for the quantum number ($n'l'$). The shape of the function $t_{l}^{0}(\theta
,\varphi)$ in Eq. (\ref{eq:dntlm}) is given in Fig. \ref{fig:pos_dis_tlm}(c). 

The particular cases are as follows:
\begin{align}
\Delta n_{6p}^{(2)} =&\frac{1}{5}\left[  n_{6p,z}-\frac{1}{2}(n_{6p,x}%
+n_{6p,y})\right]  ,\\
\Delta n_{5d}^{(2)} =&\frac{1}{7}\left[  n_{5d,z^{2}}+\frac{1}{2}%
  (n_{5d,xz}+n_{5d,yz})\right.\nonumber\\
  &\quad\left.\vphantom{\frac{1}{7}}-(n_{5d,x^{2}-y^{2}}+n_{5d,xy})\right]  ,\\
\Delta n_{5d}^{(4)} =&\frac{1}{28}\left[  n_{5d,z^{2}}-\frac{2}{3}%
  (n_{5d,xz}+n_{5d,yz})\right.\nonumber\\
  &\qquad\left.+\frac{1}{6}(n_{5d,x^{2}-y^{2}}+n_{5d,xy})\right]
,\label{eq:dnd4}%
\end{align}
where $n_{n'l',m'}$ is the occupation number of the
($n'l'$, $m'$) orbital. We note that $\Delta n_{6p}^{(4)}=0$.
Valence contribution of $A_{2}^{0}\langle r^{2}\rangle$ and $A_{4}^{0}\langle
r^{4}\rangle$ are determined as \cite{Richter1,Hummler0},
\begin{align}
A_{2}^{0}\langle r^{2}\rangle(val) &  =F^{(2)}(4f,6p)\Delta n_{6p}^{(2)}%
+F^{(2)}(4f,5d)\Delta n_{5d}^{(2)},\label{A20val}\\
A_{4}^{0}\langle r^{4}\rangle(val) &  =F^{(4)}(4f,5d)\Delta n_{5d}^{(4)},%
\label{A40val}%
\end{align}
with the Slater-Condon parameters:
\begin{align}
  F&^{(l)}(4f,n'l')\nonumber\\
  &=\frac{e^{2}}{4\pi\varepsilon_0}\iint_0^{r_c}\frac{r_{<}^{l}%
}{r_{>}^{l+1}}r^{2}|R_{4f}(r)|^{2}r'^{2}|R_{n'l'}(r')|^{2}dr'dr > 0,
\end{align}
where $r_{<}=\min(r,r')$ and $r_{>}=\max(r,r')$. Via Eqs. (\ref{A20val}) and
(\ref{A40val}), the distribution parameters $\Delta n_{n'l'}^{(l)}$
determine $A_{l}^{0}\langle r^{l}\rangle(val)$. It may be noted that no $6p$
and $5d$ orbitals exist for $A_{6}^{0}\langle r^6\rangle(val)$. 
%%%%%%%%%%%%%%%%%%%%%%%%%%%%%%%%%%%%%%%%%%%%%%%%%%%%%%
%%%%%%%%%%%%%%%%%%%%%%%%%%%%%%%%%%%%%%%%%%%%%%%%%%%%%%%

A simple explanation for the appearance of the uniaxial MA in a Sm ion
surrounded by Fe atoms is given as follows.
Fig. \ref{fig:pos_dis_tlm}(a) shows the lattice strucutre of SmFe$_{12}$ \cite{Hirayama2,Harashima}.
Left panel of Fig. \ref{fig:pos_dis_tlm}(b) shows the location of Sm and Fe on (010) plane of the lattice.
Because of the short atomic distance between Sm and
the first nearest neighbor (n.n.) Fe($8i$) sites, the
distribution of valence electrons on Sm extends within the $a-b$ plane as
shown in Fig. \ref{fig:pos_dis_tlm}(c). According to the negative sign of
$t_{2}^{0}(\theta,\varphi)$ in Fig. \ref{fig:pos_dis_tlm}(d),
the distribution parameters defined by Eq. (\ref{eq:dntlm}) in terms of
electron numbers of $6p$ and $5d$-orbitals are negative; $\Delta
n_{6p}^{(2)}=-0.0012$, $\Delta n_{5d}^{(2)}=-0.0011$. Therefore, we obtain $A_{2}%
^{0}\langle r^{2}\rangle(val)<0$ by Eq. (\ref{A20val}) in agreement with the
numerical value of $A_{2}^{0}\langle r^{2}\rangle$ shown in TABLE
\ref{table:CFP}. As shown by Eq. (\ref{eq:K1an})
the main contribution of the MA
constant $k_1(T)$ is given by a product of $A_{2}^{0}$ and the positive value
of Stevens factor $\Theta_{2}^{0}$, and $k_1(T)$ becomes positive.
This means that the $K_1(T) >0$ because $K_1^{TM}(T)>0$.

On the other hand, second neighbor Fe(8$j$) and third neighbor
Fe(8$f$) atoms of Sm atom are situated obliquely upward as shown in Fig.
\ref{fig:pos_dis_tlm}(b). According to the negative sign of $t_{4}^{0}%
(\theta,\varphi)$ shown in Fig. \ref{fig:pos_dis_tlm}(d),
we obtained $\Delta n_{5d}^{(4)}=-0.0013$ using Eq. (\ref{eq:dnd4}), and $A_4^0
\langle r^4\rangle(val)<0$
from Eq. (\ref{A40val}).
Again the negative value is
consistent with the numerical values of $A_{4}^{0}\langle r^{4}\rangle$.
The main contribution of MA constant $k_{2}(T)$ comes from a product of
$A_{4}^{0}\langle r^{4}\rangle$ and the positive value of $\Theta_{4}^{0}$,
and results in $K_{2}(T)<0$. 

Thus, the sign of MA constants $K_{1}(T)$ and $K_{2}(T)$ are determined by the
configuration of Sm and Fe atoms in the lattice.
  In the following,
  we investigate the $J$-mixing effect on single Sm magnetic properties at $T=0$ K.

\subsection{$J$-Mixing Effect and Zero-Temperature Magnetic Properties of SmFe$_{12}$ Compound}

To clarify the $J$-mixing effect on single-ion magnetic properties,
we show the calculated results of the magnetic moments $m_{L,S}(0)$
and the MA constants $k_{1,2}(0)$ for model A, B, and C
in TABLE \ref{table:KimomABC}. 
We used Eqs. (\ref{eq:momLan}) and (\ref{eq:momSan}) for
$m_{L,S}(0)$ and Eqs. (\ref{eq:K1an}) and (\ref{eq:K2an})
for $k_{1,2}(0)$,
and the values of $A_l^m\langle r^l\rangle$, $B_{\rm ex}(0)$, and $M^{TM}(0)$
in TABLE \ref{table:CFP}. 
As a reference, we also show the results
obtained by the statistical method:
$m_{L,S}(0,{\bm 0})$ in Eqs. (\ref{eq:momnumL}) and (\ref{eq:momnumS}) and
$k_{1,2}(0)$ defined in Eqs. (\ref{eq:K1num}) and (\ref{eq:K2num}).
Both the analytical and statistical
results give $k_1(0)>0$ and $k_2(0)<0$ for three models A, B, and C.
The calculated results in model C (present model) agree best with the
statistical ones.

We find that the absolute values of $m_{L,S}(0)$ and $k_{1,2}(0)$
in model B and C are larger than those in model A,
which is attributed to inclusion of the $J$-mixing effects.
The model B proposed in the previous studies \cite{Kuzminmix,Magnani}
over-estimated the $J$-mixing effects by $1/R_J$ compared with model C,
where $R_J=0.44$ in Eq. (\ref{eq:RJ}) for SmFe$_{12}$ compound.
Actually, values of $m_{L,S}^{\rm C}$ and $k_{1,2}^{\rm C}$ in TABLE \ref{table:KimomABC}
satisfy the relation in Eq. (\ref{eq:momRJ}) and (\ref{eq:kRJ}).
The results in present study ($X$=C) quantitatively agree well with statistical ones
except for $k^{\rm C}_2(0)$.
The discrepancy in $k^{\rm C}_2(0)$ may be due to omitting the 2nd order terms of
$A_2^0\langle r^2\rangle$ in Eq. (\ref{eq:K2an}),
which have a positive contribution independent of the sign of $A_2^0\langle r^2\rangle$ \cite{Kuzmin2nd}.
%%%%%%%%%%%%%%%%%%%%%%%%%%%%%%%%%%%%%%%%%%%%%%%%%%%%%%
\begin{table}[ptb]
  \caption{
    Magnetic moments $m_{L,S}(0)$ [$\mu_{\rm B}$]
    in Eqs. (\ref{eq:momLan}) and (\ref{eq:momSan})
    and
    MA constants $k_{1,2}(0)$ [K] in Eqs. (\ref{eq:K1an}) and (\ref{eq:K2an})
    for model A, B, and C at 0 K for single Sm ion.
    Results obtained by Boltzamann statistics of
    $m_{L,S}(0,{\bm 0})$ defined by Eqs. (\ref{eq:momnumL}) and (\ref{eq:momnumS}) and
    $k_{1,2}(0)$ defined by Eqs. (\ref{eq:K1num}) and (\ref{eq:K2num})
    are also shown in the fifth column.}%
\begin{ruledtabular}
\begin{tabular}{ccccc}
model&A&B&C& statistics\\ \hline
$m_L$&4.29  & 5.04  & 4.62  & 4.70\\
$m_S$&-3.57 & -5.08 & -4.24 & -4.39\\
$k_1$&60.2  & 144.5 & 97.7  & 101.1\\
$k_2$&-14.0 & -74.6 & -40.9 & -23.5
\end{tabular}
\end{ruledtabular}
\label{table:KimomABC}%
\end{table}
%%%%%%%%%%%%%%%%%%%%%%%%%%%%%%%%%%%%%%%%%%%%%%%%%%%%%%%%%%%%

\subsection{Finite Temperature Magnetic Properties \\of SmFe$_{12}$ Compound}

%%%%%%%%%%%%%%%%%%%%%%%%%%%%%%%%%%%%%%%%%%%%%%%%%%%%%%%%%%%%%%%%%%%%%%%%%%%%%%
\begin{figure}[htbp]
\begin{center}
\includegraphics[width=8.0cm]{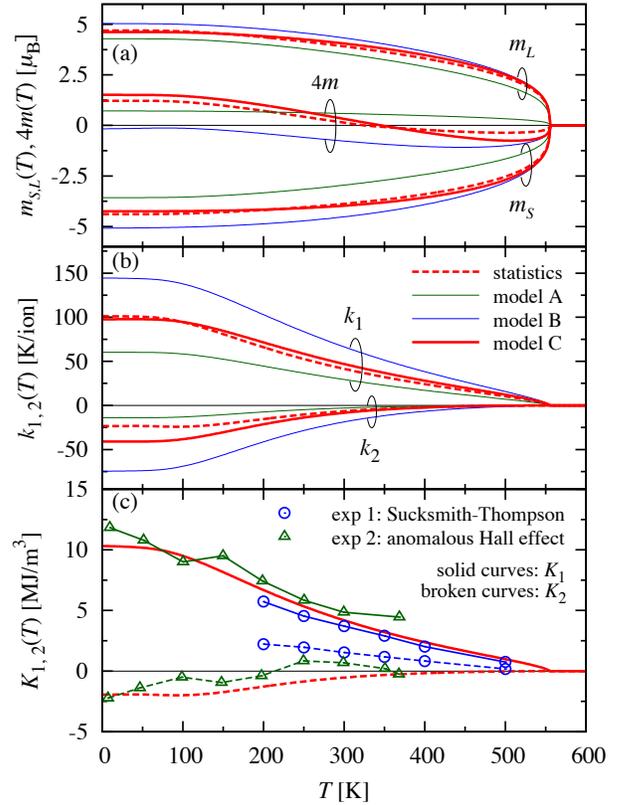}
\end{center}
\par
\caption{(color online) Temperature dependence of
  (a) magnetic moments
  of Sm ion $m_{L,S}(T)$ and $m(T)$ at ${\bm B}={\bm 0}$ and
  (b) MA constants per single Sm ion $k_{1,2}(T)$ calculated
  by using models A, B and C.
  Results obtained by Boltzmann statistics are shown by broken curves.
  (c) Temperature dependent MA constants 
  $K_{1,2}(T)$ in SmFe$_{12}$ compound 
 by using statistical method for Sm sublattice contribution $k_{1,2}(T)$,
which are compared with experimental ones by the Sucksmith-Thompson (circles)
\cite{Hirayama2} and the anomalous Hall effect (triangles) \cite{Ogawa2}.
For both calculated and experimental results in Fig. (c),
  $K_{1}(T)$ and $K_{2}(T)$ are shown by solid and broken curves, respectively.}%
\label{fig:momKi}%
\end{figure}
%%%%%%%%%%%%%%%%%%%%%%%%%%%%%%%%%%%%%%%%%%%%%%%%%%%%%%%%%%%%%%%%%%%%%%%%%%%%%%

Calculated results of finite temperature magnetic properties for a single Sm ion
in equilibrium at ${\bm e}^{TM}_0={\bm n}_c$:
the magnetic moment $m_{L,S}(T)$ in Eqs. (\ref{eq:momLan}) and (\ref{eq:momSan})
and
the MA constants $k_{1,2}(T)$ in Eqs. (\ref{eq:K1an}) and (\ref{eq:K2an})
are shown in Fig. \ref{fig:momKi}(a) and (b), respectively.
The results show that
the $J$-mixing effect in model B increases the absolute values
of both $m_{L,S}(T)$ and $k_{1,2}(T)$.
The over-estimation in model B is modified by the present model C in the
whole temperature range.
Obtained results of model C reproduce well the statistical results
for $m_{L,S}(T,{\bm 0})$ in Eqs. (\ref{eq:momnumL}) and (\ref{eq:momnumS}) and
for $k_{1,2}(T)$ in Eqs. (\ref{eq:K1num}) and (\ref{eq:K2num}) as
shown by broken lines in Fig. \ref{fig:momKi}(a) and (b).

The physical meaning of the increment of the absolute value of
$m_{L,S}(T)$ and $k_{1,2}(T)$ by $J$-mixing may be given as follows.
The expression of the free energy given by Eq. (\ref{eq:FRexan})
includes the $J$-mixing effect in the second term of the square bracket.
The term decrease $f_{\rm ex}(x)$ by $-\mu_{\rm B}B_{\rm ex}(T)\delta S(T)$,
where $\displaystyle \delta S(T)=\frac{2(L+1)}{3(J+1)}T_J^{1}(T)>0$.
Because of the decrease in $f_{\rm ex}(x)$,
the absolute value of the
spin $\langle C^{(1)}_0(\hat{\bm S})\rangle_0$ and
orbital moments $\langle C^{(1)}_0(\hat{\bm L})\rangle_0$
along ${\bm e}_0^{TM}$ are increased by $\delta S(T)$.
The tensor operators $\langle C^{(l)}_0(\hat{\bm L})\rangle_0$
for even $l$ are also increased by $\displaystyle \frac{l(l+1)}{2l+1}T_J^l(x)$,
which contribute to increase in the absolute value of
the MA constants $k_{p}^{(q)}(T)$.

The magnetic moment of Sm ion $m(T)$ is reversed at around
$T_{\rm comp}=350$ K in model C and calculation by Boltzmann statistics.
The temperature is called compensation temperature.
This phenomenon is observed also in other Sm compounds \cite{Adachi1,Adachi2}.
Zhao $et$ $al$ pointed out that
this phenomenon also appears at $T=337$ K in Sm$_2$Fe$_{17}$N$_x$
using statistical method including similar parameter values with ours
such as $\mu_{B}B_{\rm ex}(0)/k_{\rm B}=300$ K and $\lambda/k_{\rm B}=411$ K \cite{Zhao}.
Their results are comparable with ours, however,
the mechanism has not been surveyed.
In the present model C, the magnetic moment of Sm ion
can be written as
$m(T)=g_J\mu_{\rm B}JB_J^1(x)-\mu_{\rm B}\delta S(T)$.
Because $\mu_{\rm B}\delta S(T)$ is proportional to $T_J^{1,{\rm C}}(x)$ and
monotonically increasing with temperature below $T/T_{\rm C}=0.8$ as shown
in Fig. \ref{fig:Tlm_scaled}(a),
the term compensates the $g_J\mu_{\rm B}JB_J^1(x)$ at $T_{\rm comp}$.

Fig. \ref{fig:momKi}(c) shows the results of $K_{1}(T)$ and $K_{2}(T)$ 
obtained by statistical method in SmFe$_{12}$ compound, which are
compared with experimental ones denoted by exp 1 and exp 2 measured by the
Sucksmith-Thompson method \cite{Hirayama2} and anomalous Hall effect
\cite{Ogawa2}, respectively. At the whole temperature region the results of
$K_{1}(T)$ agree well with the experiments. Our statistical results qualitatively
reproduce the experimental results below 200 K. The negative $K_{2}(T)$ at low
temperatures is origin of first-order magnetization process (FOMP) as discussed below.

\subsection{Thermodynamic Properties \\of SmFe$_{12}$ Compound}

%%%%%%%%%%%%%%%%%%%%%%%%%%%%%%%%%%%%%%%%%%%%%%%%%%%%%%%%%%%%%%%%%%%%%%%%%%%%%%
\begin{figure}[htbp]
\begin{center}
\includegraphics[width=8.0cm]{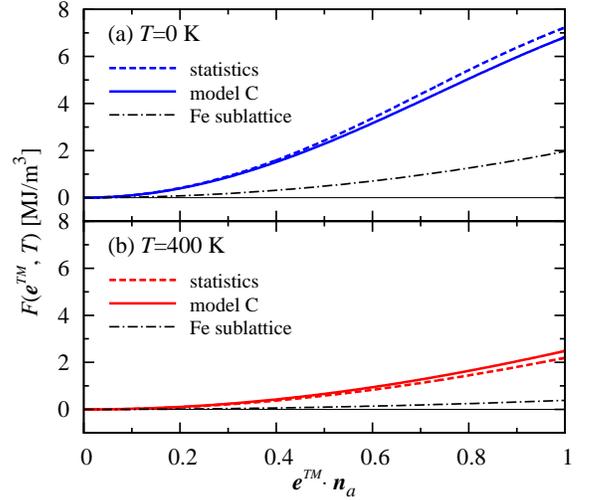}
\end{center}
\par
\caption{(color online)
Dependence of Helmholtz free energy density
on ${\bm e}^{TM}\cdot{\bm n}_a$ with ${\bm e}^{TM}\cdot{\bm n}_b=0$ in SmFe$_{12}$ at
(a) $T=0$ and (b) 400 K.
Analytical results $F({\bm e}^{TM},T)$ in Eq. (\ref{eq:Fan})
and results of $G({\bm e}^{TM},T,{\bm 0})$ obtained by Boltzmann statistics
in Eq. (\ref{eq:Gnum})
are shown by solid and broken curves, respectively,
in which the contribution from Fe sublattice is included.
The dashed-dotted curves represent the Fe sublattice MA energy:
$K_1^{TM}(T)({\bm e}^{TM}\cdot{\bm n}_a)^2$}%
\label{fig:Helmholtz}%
\end{figure}
%%%%%%%%%%%%%%%%%%%%%%%%%%%%%%%%%%%%%%%%%%%%%%%%%%%%%%%%%%%%%%%%%%%%%%%%%%%%%%

Fig. \ref{fig:Helmholtz} shows calculated results of the Helmholtz free energy density
$F({\bm e}^{TM},T)$ given in Eq. (\ref{eq:Fan}) for model C as a function of
${\bm e}^{TM}\cdot{\bm n}_a$ with ${\bm e}^{TM}\cdot{\bm n}_b=0$ at $T=0$ K
and $400$ K, where ${\bm n}_{a(b)}$ is unit vector parallel to $a(b)$-axis.
The results are compared with statistical ones
of $G({\bm e}^{TM},T,{\bm 0})$ in Eq. (\ref{eq:Gnum}).
When the direction of ${\bm e}^{TM}$ is changed,
the free energy density on both Sm and Fe sublattice are increased.
  For the Sm sublattice, the energy increase originates from the
CF, which can be expressed
by the $\sum_jf_{{\rm CF},j}({\bm e}^{TM},T)$ in Eq. (\ref{eq:FRCFan2}),
and for Fe sublattice, the energy increase can be written by:
$K_1^{TM}(T)\sin^2\theta^{TM}$ with $K_1^{TM}(T)=1.966$ and $0.387$ MJ/m$^3$
at 0 and 400 K, respectively,
which are much smaller than those of Sm sublattice
$\sum_jk_{1,j}(T)/V_0=8.059$ and $2.310$ MJ/m$^3$.
The analytical results agree well with statistical ones.

%%%%%%%%%%%%%%%%%%%%%%%%%%%%%%%%%%%%%%%%%%%%%%%%%%%%%%%%%%%%%%%%%%%%%%%%%%%%%%
\begin{figure}[htbp]
\begin{center}
\includegraphics[width=8.0cm]{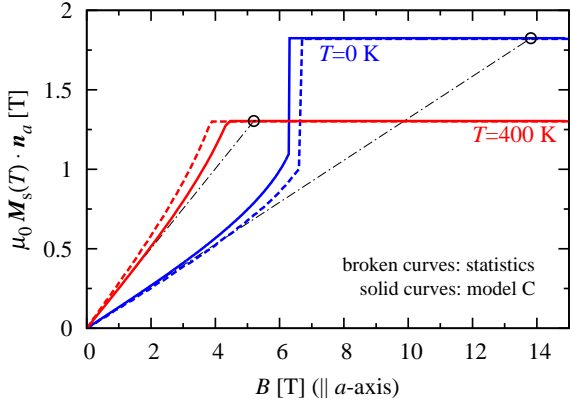}
\end{center}
\caption{(color online) Magnetization curves of SmFe$_{12}$ at $T=0$
and 400 K with applied field $B$ parallel to $a$-axis in the equilibrium
calculated by analytical (solid curves) and
statistical (broken curves) methods
in Eqs. (\ref{eq:Magan}) and (\ref{eq:Magnum}), respectively, where $\mu_0$ is
the magnetic constant.
Dashed-dotted lines show tangent lines of magnetization curves at $B=0$:
$y=[\mu_0M_{\rm s}(T)^2/2K_1] B$.
Values of $B$ at the circles correspond to the nucleation field $B_{\rm N}(T)$
obtained by using the free energy density of model C (see text).}
\label{fig:maganSmFe12}%
\end{figure}
%%%%%%%%%%%%%%%%%%%%%%%%%%%%%%%%%%%%%%%%%%%%%%%%%%%%%%%%%%%%%%%%%%%%%%%%%%%%%%

Fig. \ref{fig:maganSmFe12} shows calculated results of magnetization curves
in the equilibrium states of SmFe$_{12}$ at $T=0$ and 400 K,
where the magnetic field ${\bm B}$ is applied along $a$-axis.
Analytical results of the magnetization along the $a$-axis
are compared with statistical ones.
We have confirmed that the results in model C well reproduce the statistical ones.
At $T=0$,
we find characteristic behavior of an abrupt change in the magnetization
${\bm M}_{\rm s}(T)\cdot{\bm n}_a$ at $B=B_{\rm FP}$.
The change is called first-order magnetization process (FOMP)
  and the $B_{\rm FP}$ is called as FOMP field.
At $T$=400 K, no FOMP appears in both analytical and statistical results and
the magnetization saturates at the MA field $B_{\rm A}$.
In SmFe$_{12}$, the magnetization curve at low temperatures were not reported,
however, in SmFe$_{11}$Ti compound, FOMP observed at $T=5$ K and
$B_{\rm FP}=10$ T \cite{Hu},
which is qualitatively consistent with our results.

%%%%%%%%%%%%%%%%%%%%%%%%%%%%%%%%%%%%%%%%%%%%%%%%%%%%%%%%%%%%%%%%%%%%%%%%%%%%%%
\begin{figure}[htbp]
\begin{center}
\includegraphics[width=8.0cm]{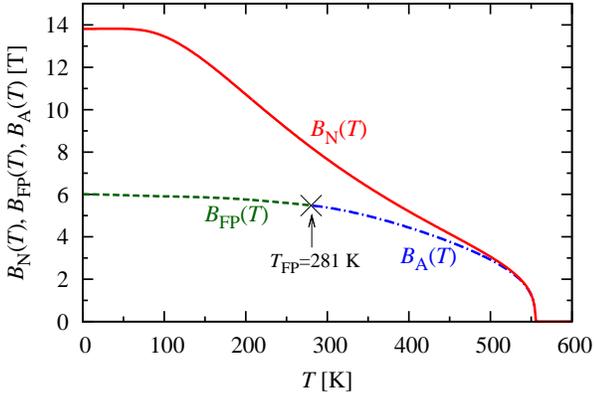}
\end{center}
\par
\caption{(color online) 
  Temperature dependence of the nucleation field $B_{\rm N}(T)$,
  the MA field $B_{\rm A}(T)$, and the
first-order magnetization process (FOMP) field $B_{\rm FP}(T)$
obtained by using the approximated free energy density
of model C neglecting  $K_3(T)$, $K_2^{1}(T)$, and $K_3^{1}(T)$,
the details of which are shown in appendix \ref{sec:FOMP}.
$B_{\rm FP}(T)$ is the point of discontinuity in the
FOMP realized below FOMP temperature $T_{\rm FP}$ (see text).}
\label{fig:BNBFPBA}%
\end{figure}
%%%%%%%%%%%%%%%%%%%%%%%%%%%%%%%%%%%%%%%%%%%%%%%%%%%%%%%%%%%%%%%%%%%%%%%%%%%%%%

Let us consider the magnetization process along $c$-axis and
estimate nucleation field $B_{\rm N}$ in the model C.
The magnetization is first saturated as $M_{\rm s}(T){\bm n}_c$ along $c$-axis
by an infinitesimal field.
Then the direction of the magnetic field is reversed and the magnitude is
increased as $-B{\bm n}_c$.
The original state continues to exist as a quasi-stable state as far as
the condition for a first-order variation $\delta G({\bm n}_c,T,-B{\bm n}_c)>0$
is satisfied. The magnetization tends to decline when $\delta G(nc,T,-B{\bm n}_c)=0$.
The applied magnetic field at which the latter condition is satisfied is the nucleation field,
which has been given as $B_{\rm N}=2K_1(T)/M_{\rm s}(T)$ \cite{Kronmuller1}.

Because $B_{\rm N}$ corresponds to the field at which the magnetization begins
to decline with infinitesimally angle $\theta$ against $c$-axis,
the magnitude $B_{\rm N}$ in realistic system can be estimated once
the magnetization curve is obtained along a hard axis.
Fig. \ref{fig:maganSmFe12} shows the magnetization curve along $a$-axis calculated in the model C.
$B_{\rm N}$ is given by a crossing point of the magnetization curve in the saturated state
and the tangential line of the magnetization curve at zero field
$y=\left[\mu_0M_{\rm s}(T)^2/2K_1(T)\right]$.
When the value of $y=\mu_0 M_{\rm s}$, the magnetic field coincides with $B_{\rm N}$ defined in Eq. (\ref{eq:BN}). 

  The magnetization curves along hard and easy-axis
  in the case of $K_1(T) > 0$ and $K_2(T) < 0$
  can be characterized by the nucleation field $B_{\rm N}(T)$, the FOMP field $B_{\rm FP}(T)$,
  and the MA field $B_{\rm A}(T)$.
  These values are analytically expressed by using the ratio $\gamma(T)=K_1(T)/K_2(T)$ as,
\begin{align}
  B_{\rm N}(T)&=\frac{2K_1(T)}{M_{\rm s}(T)},\\
  B_{\rm FP}(T)&=B_{\rm N}[x_{\rm FP}(T)+2\gamma(T)x_{\rm FP}(T)^3]\label{eq:FPap}\\
  &\qquad\qquad\qquad\qquad(0 < x_{\rm FP}(T)<1),\nonumber\\
  B_{\rm A}(T)&=B_{\rm N}[1+2\gamma(T)]\\
  &\qquad\qquad\qquad\qquad(x_{\rm FP}(T)>1),\nonumber
\end{align}
with
\begin{align}
  x_{\rm FP}(T)&=\frac{1}{3}\left(  -1+\sqrt{-\frac{3}{\gamma(T)}-2}\right),
\end{align}
where we use the approximate free energy density:
  $F({\bm e}^{TM},T)=K_1(T)\sin^2\theta^{TM}
  +K_2(T)\sin^4\theta^{TM}$, in which the small contributions
  $K_3(T)$, $K_{2,3}^{1}(T)$ are neglected.
  Details are shown in Appendix \ref{sec:FOMP}.
  Calculated results of $B_{\rm N}(T)$, $B_{\rm FP}(T)$, and $B_{\rm A}(T)$
  are shown in Fig. \ref{fig:BNBFPBA}.
The condition of FOMP appearance in model C is given by
  $-K_2(T) < K_1(T) < -6K_2(T)$ between $0<x_{\rm FP}(T)<1$ in Eq. (\ref{eq:FPap}).
As for SmFe$_{12}$ compound, the FOMP is realized below $T=281$ K $\equiv$ $T_{\rm FP}$, 
    which is analytically obtained from the condition: $K_1(T)=-6K_2(T)$.
The curves of $B_{\rm FP}(T)$ and $B_{\rm A}(T)$ are continuously connected at
$T_{\rm FP}$, which is called as FOMP temperature.
  When $K_1(T)<-K_2(T)$ the magnetization direction is in-plane at ${\bm B}={\bm 0}$.

\section{Summary}

The temperature dependence of magnetic anisotropy (MA) constants and magnetization of SmFe$_{12}$
were investigated by using two methods for the model Hamiltonian
which combines quantum and phenomenological ones for rare-earth ($R$) and Fe subsystem,
respectively.
Parameter values of $R$ Hamiltonian were determined by the first-principles.
First method adopts a numerical
procedure with Boltzmann statistics for the Sm $4f$ electrons.
The other one is an analytical method which deals
with the magnetic states of $R$ ions with strong
mixing of states with different quantum number of angular momentum $J$ ($J$-mixing).
We have modified the previous analytical methods for Sm ions which
have relatively small spin-orbit interaction,
and clarified that they over-estimate the $J$-mixing effects
for Sm-transition metal compounds.
It has been shown that the
results of our analytical method agree with those obtained by statistical method.
Our analytical method revealed that
the increasing spin angular momentum with $J$-mixing caused by strong exchange field,
enhances the absolute value of orbital angular momentum and MA constants
via spin-orbit interaction, and that
these $J$-mixing effects remain even above room temperature.
The calculated results of MA constants show that $K_{1}(T)>0$ and $K_{2}(T)<0$ in SmFe$_{12}$
in consistent with experiment.

The peculiar temperature dependence known as first-order magnetization process
(FOMP) in SmFe$_{12}$ has been attributed to the negative $K_{2}$.
It was also
verified that the requirement for the appearance of FOMP is given as
$-K_2<K_{1}<-6K_{2}$.
The positive (negative) $K_{1(2)}$ appears due to an increase in
the crystal field parameter $A_{2}^{0}\langle r^{2}\rangle$ ($A_4^0\langle r^4\rangle$)
caused by hybridization
between $3d$-electrons of Fe on $8i$ ($8j$) site and
$5d$ and $6p$ valence electrons on Sm.
The mechanism of $K_{1}>0$ and $K_{2}<0$ in SmFe$_{12}$ has been thus clarified by
using the expressions of $K_{1}$ and $K_{2}$ obtained in the analytical
method. Shortly, the sign of $K_{1}$ and $K_{2}$ in SmFe$_{12}$ is attributed
to the characteristic lattice structure around Sm ions, that is,
crystallographic 2$b$-sites on $c$-axis adjacent to Sm are vacant. We also present
results on the magnetization process and nucleation fields by calculating
Gibbs free energy. 

The present method will be applied to derive a general expression of the free
energy to analyze MA of non-uniform systems such as disordered compounds,
surfaces, and interfaces. The results will be reported in a forthcoming paper.

\begin{acknowledgements}
This work is supported by ESICMM Grant Number 12016013 and ESICMM is funded by
Ministry of Education, Culture, Sports, Science and Technology (MEXT).
T. Y. was supported by JSPS KAKENHI Grant Numbers JP18K04678.
P. N. was supported by the project Solid21.
\end{acknowledgements}

\appendix

\section{Magnetization Process in Condition \\of $K_1(T)>0$ and $K_2(T)<0$} \label{sec:FOMP}

To investigate the magnetization process in equilibrium along the
$c$-plane (e.g. $a$-axis), we introduce
the simplified model with magnetic anisotropy constants
$K_1(T)>0$ and $K_2(T)<0$, which can be expressed by the
Gibbs free energy as:
\begin{align}
  G(x,T,B)=K_1(T)x^2+K_2(T)x^4-&BM_{\rm s}(T)x\\
  &(|x|\le 1),\nonumber
\end{align}
where
  $x={\bm M}_{\rm s}(T)\cdot{\bm n}_a/M_{\rm s}(T)$
  with total magnetization ${\bm M}_{\rm s}$ and
  unit vector parallel to $a$-axis ${\bm n}_{a}$.
  $T$ and ${\bm B}=B{\bm n}_a$ ($B>0$)
are temperature and applied magnetic field, respectively.
The equilibrium condition is:
\begin{equation}
  G(x_0,T,B)=\min_{|x|\le 1}G(x,T,B), \nonumber
\end{equation}
where $x=x_0(T,B)$ gives minimum of $G(x,T,B)$.
For $K_1(T) \le -K_2(T)$, the magnetization is always tilted to the $a$-axis
  direction due to $x_0(T, B)=1$.
  Otherwise, the magnetization curve is given by:
\begin{equation}
  {\bm M}_{\rm s}(T,B)\cdot{\bm n}_a=M_{\rm s}(T)x_0(T,B).
\end{equation}  
The first-order magnetization process (FOMP)
appears, when $x_0(T,B)$ has two values at certain $B$,
which is called as FOMP field $B_{\rm FP}$.

To determine the $x_0(T,B)$ for $K_1(T)>-K_2(T)$,
we show the first and second derivative of $G(x,T,B)$ with respect to $x$ as:
\begin{align}
\frac{\partial G(x,T,B)}{\partial x}&=2K_1(T)x+4K_2(T)x^3-BM_{\rm s}(T),\label{eq:gp}\\
\frac{\partial^2 G(x,T,B)}{\partial x^2}&=2K_1(T)+12K_2(T)x^2.
\end{align}
A inflection point of $G(x,T,B)$
for $x>0$ at fixed $T$ and $B$ is given by $x_c(T)=\sqrt{-K_1(T)/6K_2(T)}$.
%where $x_1$ depends on $B$
Hereafter, we consider following two cases:
$x_c(T)\ge 1$ and $x_c(T)<1$.

(i) The case of $x_c(T)\ge 1$

$x_0(T,B)$ is obtained from the condition $\partial G(x,T,B)/\partial x= 0$
  for $0< x\le1$,
because $\partial^2 G(x,T,B)/\partial x^2> 0$ is always satisfied.
%The magnetization along $a$-axis is
%${\bm M}_{\rm s}(T)\cdot{\bm n}_a=\min\{M_{\rm s}(T)x_1,M_{\rm s}(T)\}$.
The saturating point of magnetization $x_0(T,B)=1$ is obtained from
the condition $\partial G(x,T,B)/\partial x|_{x=1}=0$ as:
\begin{align}
  B=\frac{2K_1(T)}{M_{\rm s}(T)}[1+2\gamma(T)]\equiv B_{\rm A}(T),
\end{align}
where $\gamma(T)=K_2(T)/K_1(T)$.
The $B_{\rm A}$ is so-called anisotropy field.

(ii) The case of $x_c(T)<1$

  $x_0(T,B)$ is obtained from the condition
\begin{align}
  G(x_0,T,B)=\min\left[ G(x_e,T,B), G(1,T,B)\right],
\end{align}  
where $x_e(T,B)$ is determined by the condition of local minium as:
$\partial G(x,T,B)/\partial x=0$ and $x_e(T,B)<x_c(T)$.
  In the magnetization process, $x_0(T,B)$ is continuously increased
  from zero with increasing $B$ according to $x_0(T,B)=x_e(T,B)$ for
  $G(x_e,T,B)<G(1,T,B)$.
  At $B=B_{\rm FP}$ such that $G(x_e,T,B)=G(1,T,B)$ is satisfied,
  $x_0(T,B)$ shows the abrupt jump and becomes saturated value of $x_e(T,B)=1$.
The condition is rewritten as:
\begin{align}
  \label{eq:grad}
  (x_0-1)\left[3K_2(T)x_0^2+2K_2x_0+K_1(T)+K_2(T)\right]=0.
\end{align}
By solving the Eq. (\ref{eq:grad}) for $0<x_0\le 1$,
two minimum points of $G(x,T,B)$ with respect to $x$
are obtained at $x_0(T,B)=1$ and
\begin{equation}
x_0(T,B)=\frac{1}{3}\left(  -1+\sqrt{-\frac{3}{\gamma(T)}-2}\right)\equiv x_{\rm FP}(T).\label{eq:xFP}%
\end{equation}
By using $x_{\rm FP}(T)$, the field at which the FOMP occurs is determined by:
\begin{equation}
  B=\frac{2K_1(T)}{M_{\rm s}(T)}[x_{\rm FP}(T)+2\gamma(T)x_{\rm FP}(T)^3]
  \equiv B_{\rm FP}(T).\label{eq:HFP}
\end{equation}

As a result, for $-1<\gamma(T)< -1/6$, the FOMP occurs between ${\bm M}_{\rm s}(T)\cdot{\bm n}_a
=M_{\rm s}(T)x_{\rm FP}(T)$ and $M_{\rm s}(T)$.

\end{document}